\documentclass[a4paper,fleqn,usenatbib]{mnras}
\usepackage{longtable}
\usepackage{newtxtext,newtxmath}
\usepackage[T1]{fontenc}
\usepackage{ae,aecompl}
\usepackage{graphicx}	
\usepackage{amsmath}	
\defcitealias{ccm89}{CCM89}
\defcitealias{ngc5904}{Paper I}
\defcitealias{ngc6205}{Paper II}
\defcitealias{stetson2019}{SPZ19}
\defcitealias{zloczewski2012}{ZKR12}
\defcitealias{hargis2004}{HSB04}
\defcitealias{grundahl1999}{GCL99}
\defcitealias{bellazzini2001}{BPF01}
\defcitealias{sollima2016}{SFL16}
\defcitealias{rosenberg2000}{RPS00}
\defcitealias{bonatto2013}{BCK13}
\defcitealias{vasiliev2021}{VB21}

\title[Isochrone fitting of NGC\,288, NGC\,362, and NGC\,6218]
{Isochrone fitting of Galactic globular clusters -- III. NGC\,288, NGC\,362, and NGC\,6218 (M12)}

\author[G. A. Gontcharov et al.]{
George~A.~Gontcharov,$^{1}$\thanks{E-mail: georgegontcharov@yahoo.com}
Maxim~Yu.~Khovritchev,$^{1,2}$
Aleksandr~V.~Mosenkov,$^{1,3}$
\newauthor
Vladimir~B.~Il'in,$^{1,2,4}$
Alexander~A.~Marchuk,$^{1,2}$
Sergey~S.~Savchenko,$^{1,2,5}$
\newauthor
Anton~A.~Smirnov,$^{1,2}$
Pavel~A.~Usachev,$^{1,2,5}$
and Denis~M.~Poliakov$^{1,2}$   
\\
$^{1}$Central (Pulkovo) Astronomical Observatory, Russian Academy of Sciences, Pulkovskoye chaussee 65/1, St. Petersburg 196140, Russia\\
$^{2}$Saint Petersburg State University, Universitetskij pr. 28, St. Petersburg 198504, Russia\\
$^{3}$Department of Physics and Astronomy, N283 ESC, Brigham Young University, Provo, UT 84602, USA\\
$^{4}$Saint Petersburg University of Aerospace Instrumentation, Bol. Morskaya ul. 67A, St. Petersburg 190000, Russia\\
$^{5}$Special Astrophysical Observatory, Russian Academy of Sciences, 369167 Nizhnij Arkhyz, Russia\\
}

\date{Accepted XXX. Received YYY; in original form ZZZ}

\pubyear{2021}

\begin{document}
\label{firstpage}
\pagerange{\pageref{firstpage}--\pageref{lastpage}}
\maketitle

\begin{abstract}
We present new isochrone fits to colour--magnitude diagrams of the Galactic globular clusters NGC\,288, NGC\,362, and NGC\,6218 (M12).
We utilize a lot of photometric bands from the ultraviolet to mid-infrared by use of data from the {\it HST}, {\it Gaia}, unWISE, 
Pan-STARRS, and other photometric sources.
In our isochrone fitting we use theoretical models and isochrones from the Dartmouth Stellar Evolution Program and 
Bag of Stellar Tracks and Isochrones for $\alpha$--enhanced abundance [$\alpha$/Fe]$=+0.40$, different helium abundances, and
a metallicity of about [Fe/H]$=-1.3$ adopted from the literature.
We derive the most probable distances $8.96\pm0.05$, $8.98\pm0.06$, and $5.04\pm0.05$ kpc, 
ages $13.5\pm1.1$, $11.0\pm0.6$, and $13.8\pm1.1$ Gyr, 
extinctions $A_\mathrm{V}=0.08\pm0.03$, $0.11\pm0.04$, and $0.63\pm0.03$ mag,
and reddenings $E(B-V)=0.014\pm0.010$, $0.028\pm0.011$, and $0.189\pm0.010$ mag
for NGC\,288, NGC\,362, and NGC\,6218, respectively.
The distance estimates from the different models are consistent, while those of age, extinction, and reddening are not.
The uncertainties of age, extinction, and reddening are dominated by some intrinsic systematic differences between the models.
However, the models agree in their relative age estimates: 
NGC\,362 is $2.6\pm0.5$ Gyr younger than NGC\,288 and $2.8\pm0.5$ Gyr younger than NGC\,6218, confirming age as the second parameter for these clusters.
We provide reliable lists of the cluster members and precise cluster proper motions from the {\it Gaia} Early Data Release 3.
\end{abstract}

\begin{keywords}
Hertzsprung--Russell and colour--magnitude diagrams --
dust, extinction --
globular clusters: general --
globular clusters: individual: NGC\,288, NGC\,362, NGC\,6218 (M12)
\end{keywords}

\section{Introduction}
\label{intro}

In \citet[][hereafter Paper I]{ngc5904} and \citet[][hereafter Paper II]{ngc6205} we presented isochrone fits to colour--magnitude diagrams 
(CMDs) of the Galactic globular clusters (GCs) NGC\,5904 (M5) and NGC\,6205 (M13). 
We utilized accurate photometry of individual stars in many ultraviolet (UV), optical, and infrared (IR) bands by use of datasets from 
the {\it Hubble Space Telescope (HST)}, {\it Gaia} Data Release 2 (DR2, \citealt{gaiaevans}),
{\it Wide-field Infrared Survey Explorer (WISE}, \citealt{wise}), 
Panoramic Survey Telescope and Rapid Response System Data Release I (Pan-STARRS, PS1, \citealt{bernard2014,chambers2016}), and other photometric sources. 
In our isochrone fitting we used
the Dartmouth Stellar Evolution Program (DSEP, \citealt{dotter2007})\footnote{\url{http://stellar.dartmouth.edu/models/}},
a Bag of Stellar Tracks and Isochrones (BaSTI-IAC, \citealt{newbasti})\footnote{\url{http://basti-iac.oa-abruzzo.inaf.it/index.html}}, and other
theoretical stellar evolution models and related isochrones.
We employed the models for both the solar-scaled and $\alpha$--helium--enhanced abundances, with spectroscopic metallicities adopted from the literature.

All or almost all GCs contain multiple stellar populations \citep{monelli2013}.
In spite of this, a dominant population or a mix of populations can be accurately fitted in some CMDs of some GCs.
NGC\,5904 and NGC\,6205 are examples of such GCs.
In each CMD, except for some in the UV, we derived age, distance, and reddening for a dominant population or a mix of populations by isochrone fitting 
for different stages of stellar evolution, which can be well recognized in the CMDs. 
They are the main sequence (MS), its turn-off (TO), the subgiant branch (SGB), red giant branch (RGB), horizontal branch (HB), and asymptotic giant 
branch (AGB).
Moreover, the models underlying these isochrones were verified by this fitting.

The derived distances and ages are different for the UV, optical, and IR photometry used, while the derived ages and reddenings are different 
for the different models.
However, we obtained convergent estimates of the most probable distances and extinctions in all the bands, but less consistent estimates 
of ages.
The extinctions appear to be twice as high as generally accepted for these GCs.
We determined empirical extinction laws (dependence of extinction on wavelength), which agree with the law of \citet[][hereafter CCM89]{ccm89} with the 
best-fitting extinction-to-reddening ratio $R_\mathrm{V}\equiv A_\mathrm{V}/E(B-V)=3.6\pm0.05$ and $3.1^{+1.6}_{-1.1}$ for NGC\,5904 and NGC\,6205, 
respectively.

The aim of this paper is to expand our research of Galactic GCs by exploiting a multiband photometry and up-to-date theoretical stellar evolution models
for the other three GCs NGC\,288, NGC\,362, and NGC\,6218 (M12), despite the presence of multiple populations in these clusters 
\citep{monelli2013,milone2017}.
Following \citetalias{ngc5904} and \citetalias{ngc6205}, we adopt the spectroscopic metallicity and $\alpha$--helium--enhancement from the literature in 
order to obtain the most probable ages, distances, and empirical extinction laws for these GCs. 
Also, we estimate the accuracy and consistency of the models/isochrones under consideration.

As this is the third paper in this series devoted to investigation of GCs, the full details of our analysis, which we carry out in this study, 
are given in our \citetalias{ngc5904} and \citetalias{ngc6205}. 
We refer the interested reader to those papers, especially to appendix A of  \citetalias{ngc6205}, although throughout this paper we briefly describe 
some crucial points of those studies.

This paper is organized as follows. We describe some key properties of NGC\,288, NGC\,362, and NGC\,6218 in Sect.~\ref{metal}.
In Sect.~\ref{photo} we describe the photometry used, our cleaning of the datasets and creation of the fiducial sequences in the CMDs.
In Sect.~\ref{iso} we describe the theoretical models and corresponding isochrones used.
The results of our isochrone fitting with their discussion are presented in Sect.~\ref{results}.
We summarize our main findings and conclusions in Sect.~\ref{conclusions}.

\begin{table*}
\def\baselinestretch{1}\normalsize\normalsize
\caption[]{Some properties of the clusters under consideration. 
A typical precision of the data is a few units of the last decimal place, unless otherwise stated.
The {\it Gaia} EDR3 median parallax is calculated by us in Sect.~\ref{edr3}.
}
\label{properties}
\[
\begin{tabular}{lccc}
\hline
\noalign{\smallskip}
 Property            &  NGC\,288  &  NGC\,362  &  NGC\,6218 \\
\hline
\noalign{\smallskip}
R.A. J2000 (h~m~s) from \citet{goldsbury2010}                     & \hphantom{$-$}00 52 45 & \hphantom{$-$}01 03 14 & \hphantom{$-$}16 47 14 \\
Decl. J2000 ($\degr$ $\arcmin$ $\arcsec$) from \citet{goldsbury2010}  & $-26$ 34 57     & $-70$ 50 56     & $-01$ 56 55 \\
Galactic longitude ($\degr$) from \citet{goldsbury2010}               & 151.2851        & 301.5330        & 15.7151 \\
Galactic latitude ($\degr$) from \citet{goldsbury2010}                & $-89.3804$      & $-46.2474$      & +26.3134 \\
Angular diameter (arcmin) from \citet{bica2019}                       & 15             & 15             & 19 \\
Distance from the Sun (kpc) from \citet{harris}, 2010 revision\footnotemark\     & 8.9             & 8.6             & 4.8 \\
Distance from the Sun (kpc) from \citet{baumgardt2021}                           & $8.99\pm0.09$   & $8.83\pm0.10$   & $5.11\pm0.05$ \\
{\it Gaia} EDR3 median parallax (mas)                                 & $0.114\pm0.011$ & $0.119\pm0.011$ & $0.210\pm0.011$ \\
$[$Fe$/$H$]$ from \citet{carretta2009}                                & $-1.32\pm0.02$  & $-1.30\pm0.04$  & $-1.33\pm0.02$ \\
$[$Fe$/$H$]$ from \citet{marsakov2019}                                & $-1.37\pm0.04$  & $-1.20\pm0.08$  & $-1.35\pm0.05$ \\
$[\alpha/$Fe$]$ from \citet{carretta2010}                             & 0.42            & 0.30            & 0.41 \\
Mean differential reddening $\overline{\delta E(B-V)}$ (mag) from \citetalias{bonatto2013} & $0.047\pm0.018$ & $0.032\pm0.009$ & $0.027\pm0.008$ \\
Maximum differential reddening $\delta E(B-V)_\mathrm{max}$ (mag) from \citetalias{bonatto2013}    & 0.091           & 0.056           & 0.056 \\
\noalign{\smallskip}
$A_\mathrm{V}$ (mag) from \citet{harris}, 2010 revision         & 0.09     & 0.16    & 0.59 \\
$A_\mathrm{V}$ (mag) from \citet{sfd98}                         & 0.04     & 0.10    & 0.54 \\
$A_\mathrm{V}$ (mag) from \citet{schlaflyfinkbeiner2011}        & 0.03     & 0.09    & 0.47 \\
$A_\mathrm{V}$ (mag) from \citet{2015ApJ...798...88M}           & 0.05     & 0.12    & 0.50 \\
\noalign{\smallskip}
$E(B-V)$ (mag) from \citet{harris}, 2010 revision               & 0.03     & 0.05    & 0.19 \\
$E(B-V)$ (mag) from \citet{sfd98}                               & 0.01     & 0.03    & 0.18 \\
$E(B-V)$ (mag) from \citet{schlaflyfinkbeiner2011}              & 0.01     & 0.03    & 0.15 \\
$E(B-V)$ (mag) from \citet{2015ApJ...798...88M}                 & 0.01     & 0.04    & 0.16 \\
\hline
\end{tabular}
\]
\end{table*}
\footnotetext{The commonly used database of GCs by \citet{harris} (\url{https://www.physics.mcmaster.ca/~harris/mwgc.dat}), 2010 revision.}

\section{Properties of the clusters}
\label{metal}

We select these GCs for our study due to a rich photometric material available for the astronomical community, a rather low foreground and 
differential reddening\footnote{In this paper by differential reddening, we mean variations of reddening over a cluster field, but see our note at 
the end of this section.},
as well as accurate spectroscopic estimates of their metallicity.
Some general properties of NGC\,288, NGC\,362, and NGC\,6218 (M12) are presented in Table~\ref{properties}.

These GCs are of particular interest since NGC\,288 is close to the South Galactic Pole, while NGC\,362 is located in the sky about 2$\degr$ 
south (in Galactic coordinates) of the optical centre of the Small Magellanic Cloud (SMC). The latter means that SMC stars represent
background for NGC\,362 and inevitably contaminate any CMD for this cluster.
This contamination should be removed, for example, by use of the proper motions (PMs) from the {\it Gaia} Early Data Release 3 (EDR3, 
\citealt{gaia2021a}), as done in Sect.~\ref{edr3}.

Moreover, NGC\,288 and NGC\,362 are a well studied second-parameter pair.
This means that, being of a similar metallicity, $\alpha$--enrichment, and even distance, these GCs have different HB morphologies:
NGC\,288 and NGC\,362 have their HBs almost completely consisting of stars on the blue and red side of the instability strip, respectively.
The HB of NGC\,6218 is rather similar to that of NGC\,288. Moreover, NGC\,5904, studied by us in \citetalias{ngc5904}, has a similar metallicity.
Hence, all four GCs should be considered as a second-parameter quartet \citep{vandenberg2013}.
A lot of studies \citep[e.g. ][]{bolte1989,lee1994,bellazzini2001,marin-franch2009} have discussed whether age is the second parameter
after metallicity which is responsible for this HB difference.
In particular, \citet{bolte1989} found an age difference as large as 5 Gyr between NGC\,288 and NGC\,362, in contrast to \citet{marin-franch2009}
who found a nearly zero age difference\footnote{`Our results indicate that NGC\,0288 and NGC\,0362 have the same age within $\pm0.9$ Gyr' 
\citep{marin-franch2009}.}.
Based on the most precise data and models, we intend to answer whether age is the second parameter for these GCs.

We adopt models with [Fe/H] close to that from \citet{carretta2009}, as indicated in Sect.~\ref{iso}.
In contrast to \citetalias{ngc5904} and \citetalias{ngc6205}, we check [Fe/H] from the isochrone fitting for the tilt of the RGB.
Brighter parts of the RGB are affected by saturation and crowding. This leads to a systematic uncertainty of about 0.1 dex in the [Fe/H] 
derived from the RGB tilt.
However, we conclude that the observed RGB tilt certainly implies $-1.4<$[Fe/H]$<-1.2$ for all the CMDs.

All these GCs have two dominating populations \citep{milone2017,lardo2018}.
Both populations are $\alpha$--enriched with $0.1<$[$\alpha$/Fe]$<0.4$ \citep{carretta2010,marsakov2019,horta2020}, but differ in helium 
abundance\footnote{`There is a broad consensus on the similarity in chemical composition between NGC\,288 and NGC\,362' \citep{bellazzini2001}.}.
It would be reasonable to assume helium content $Y\approx0.248$ and $Y\approx0.275$ for the primordial and helium-enriched populations, respectively
\citep{piotto2013,wagner2016,milone2018}.

In Table~\ref{properties} we provide both extinction $A_\mathrm{V}$ and reddening $E(B-V)$, since some of their estimates are based on a
non-\citetalias{ccm89} extinction law.
The extinction and reddening estimates from the 2D dust emission maps of \citet{sfd98}, \citet{schlaflyfinkbeiner2011}, and \citet{2015ApJ...798...88M}
are averaged over the fields of the GCs and rounded to 0.01 mag. This value nearly corresponds to the uncertainty of the mean reddening in the fields of
these GCs due to differential reddening. Some inconsistency of the extinction and reddening estimates in Table~\ref{properties} is evident. 
This will be discussed in Sect.~\ref{redext}.

Table~\ref{properties} gives the differential reddening from \citet[][hereafter BCK13]{bonatto2013}. Generally speaking, differential reddening 
should be lower than a mean reddening of a cluster.
Therefore, a rather high mean differential reddening from \citetalias{bonatto2013} $\overline{\delta E(B-V)}=0.047\pm0.018$ mag for NGC\,288, 
in particular, its maximum value $\delta E(B-V)_\mathrm{max}=0.091$ mag hardly reconciles with rather low reddening estimates $0.01<E(B-V)<0.03$ mag 
in Table~\ref{properties}.
To a lesser extent, this applies to NGC\,362.
However, the differential reddening measured by \citetalias{bonatto2013} is very hard to distinguish from the systematic variations of colours in a 
cluster field due to other reasons (hereafter field effects). Such effects include systematic photometric errors, distortion, telescope breathing, 
telescope focus change, stellar population variations, and others \citep{anderson2008}.
Generally speaking, these effects influence different photometric filters in different ways, so their observed manifestation is a slight colour shift 
of a cluster fiducial sequence as a function of location in the cluster field.
`On average this shift is zero, but the trend with position can be as large as $\pm0.02$ mag' \citep{anderson2008}.
Indeed, we found such a level of the colour variations in the field of NGC\,6205 in \citetalias{ngc6205}.
We examine differential reddening and related field effects, which appear as fiducial sequence colour variations over the cluster field, 
in Sect.~\ref{photo}.
We add some notes on these effects in Sect.~\ref{redext}.

\section{Photometry}
\label{photo}

As discussed in \citetalias{ngc5904} and \citetalias{ngc6205}, in order to derive the accurate distance, age, and reddening for a GC, we prefer using
photometric datasets covering, at least, the cluster HB, SGB, TO, and part of the RGB between the HB and SGB.

We select {\it Gaia} EDR3, Pan-STARRS, and some other initial samples/datasets of the cluster members, within the cluster diameters from
Table~\ref{properties}, by use of the VizieR and X-Match services of the Centre de Donn\'ees astronomiques de Strasbourg 
\citep{vizier}\footnote{\url{http://cds.u-strasbg.fr}}.
Then the initial datasets are cleaned, as described in Sect.~\ref{cleaning}.

The adopted cluster diameters and related truncation radii are slightly smaller than those derived by \citet{deboer2019} and in some other
recent studies.
However, for isochrone fitting, we do not need a complete sample of stars, but we do need a cleanest sample with a minimised contamination.
Adopting such truncation radii, we may lose few peripheral cluster members, but, on the other hand, we ensure a high percentage of members 
among the selected stars.

Selection of cluster members from {\it Gaia} EDR3 can be more sophisticated and precise by use of its very accurate PMs.
Therefore, unlike the other datasets in our study, we select initial {\it Gaia} EDR3 samples within initial radii which are six times 
larger than those used for the other datasets.
We use the periphery of these fields for estimating the star count surface density of the Galactic field background.
Its subtraction allows us to derive some empirical truncation radii while cleaning the initial {\it Gaia} EDR3 samples, as described 
in Sect.~\ref{edr3}.

We use the following datasets (see Table~\ref{filters}):
\begin{enumerate}
\item the {\it HST} Wide Field Camera 3 (WFC3) UV Legacy Survey of Galactic Globular Clusters (the $F275W$, $F336W$ and $F438W$ filters) and the
Wide Field Channel of the Advanced Camera for Surveys (ACS; the $F606W$ and $F814W$ filters) survey of Galactic globular clusters 
\citep{nardiello2018}\footnote{\url{http://groups.dfa.unipd.it/ESPG/treasury.php}},
\item Str\"omgren $uvby$ photometry from the 1.54 m Danish Telescope, European Southern Observatory (ESO), La Silla \citep[][hereafter GCL99]{grundahl1999},
\item a compilation of the $UBVRI$ photometry \citepalias{stetson2019}\footnote{\url{http://cdsarc.u-strasbg.fr/viz-bin/cat/J/MNRAS/485/3042}},
\item $UBVI$ photometry from the Magellanic Clouds Photometric Survey (MCPS) with the 1 m Las Campanas Swope telescope \citep{zaritsky2002},
\item photometry from \citet{piotto2002} in the $F439W$ and $F555W$ filters from the {\it HST} Wide Field and Planetary Camera 2 (WFPC2) transformed 
by the authors into the $B$ and $V$ magnitudes\footnote{We use the $B$ and $V$ instead of $F439W$ and $F555W$ filters since they are better defined 
by the models used.},
\item $BV$ photometry with the Wide Field Imager (WFI) mounted on the 2.2 m telescope, ESO, La Silla \citep[][hereafter SFL16]{sollima2016},
\item the fiducial sequence in the `$B-V$ versus $V$' plane derived by \citet{bolte1992} from the photometry with the 4 m and 0.9 m telescopes of 
the Cerro-Tololo Inter-American Observatory (CTIO),
\item $BVI$ photometry with the Kitt Peak National Observatory (KPNO) 0.9 m telescope \citep[][hereafter HSB04]{hargis2004},
\item $BV$ photometry with the 2.5 m du Pont telescope, Las Campanas \citep[][hereafter ZKR12]{zloczewski2012,kaluzny2015},
\item The Dark Energy Survey (DES) Data Release 1 photometry in the $g_\mathrm{DECam}$, $r_\mathrm{DECam}$, $i_\mathrm{DECam}$, and $z_\mathrm{DECam}$
filters obtained by \citet{des} with the Dark Energy Camera (DECam) mounted on the 4 m Blanco telescope at CTIO,
\item Parallel-Field Catalogues (the $F475W$ and $F814W$ filters) of the {\it HST} UV Legacy Survey of Galactic Globular Clusters with ACS 
\citep{simioni2018},
\item Pan-STARRS photometry in the $g_\mathrm{PS1}$, $r_\mathrm{PS1}$, $i_\mathrm{PS1}$, $z_\mathrm{PS1}$, and $y_\mathrm{PS1}$ filters 
\citep{chambers2016} and related fiducial sequences derived by \citet{bernard2014},
\item {\it Gaia} DR2 and EDR3 photometry in the $G$, $G_\mathrm{BP}$ and $G_\mathrm{RP}$ filters \citep{gaiaevans,riello2021},
\item $VI$ photometry with the 2.2 m ESO/MPI telescope, La Silla, equipped with the EFOSC2 camera, as well as the fiducial sequence 
\citep[][hereafter BPF01]{bellazzini2001},
\item $VI$ photometry with 0.91 m DUTCH telescope, ESO, La Silla \citep[][hereafter RPS00]{rosenberg2000},
\item SkyMapper Southern Sky Survey DR3 (SMSS, SMSS DR3) photometry in the $g_\mathrm{SMSS}$, $r_\mathrm{SMSS}$, $i_\mathrm{SMSS}$, and 
$z_\mathrm{SMSS}$ filters \citep{onken2019}\footnote{\url{https://skymapper.anu.edu.au}},
\item the fiducial sequence in the `$J-K$ versus $K$' plane derived by \citet{davidge1997} from the photometry with the 3.6 m Canada--France--Hawaii 
Telescope (CFHT),
\item $J$ and $K_s$ photometry obtained by \citet{cohen2015} with Infrared Side Port Imager (ISPI) mounted on the 4 m Blanco telescope at CTIO and
calibrated by use of the Two Micron All-Sky Survey (2MASS, \citealt{2mass}) stars,
\item {\it Wide-field Infrared Survey Explorer (WISE)} photometry in the $W1$ filter from the unWISE catalogue \citep{unwise},
\item photometry in the 3.6-$\mu$m filter of the {\it Spitzer} Space Telescope Infrared Array Camera (IRAC) obtained by \citet{sage} within the 
Surveying the Agents of Galaxy Evolution in the Tidally-Disrupted, Low-Metallicity Small Magellanic Cloud 
(SAGE-SMC, SAGE)\footnote{\url{http://sage.stsci.edu/}}.
\end{enumerate}

Each star has a photometry in some but not all filters.
Totally 33, 26 and 26 filters are exploited for NGC\,288, NGC\,362, and NGC\,6218, respectively.
They span a wide wavelength range between the UV and middle IR.
For each filter, Table~\ref{filters} presents the effective wavelength $\lambda_\mathrm{eff}$ in nm, number of stars and the median 
photometric precision after rejecting the poor photometry. The rejection procedure is described in Sect.~\ref{cleaning} and \ref{edr3}. 
The original datasets typically contain many more stars.
The median precision is calculated from the precisions stated by the authors of the datasets used.
The median precision is used for calculating predicted uncertainties of the derived age, distance and reddening, as described 
in appendix A of \citetalias{ngc6205} and in Sect.~\ref{results}.

\begin{table*}
\def\baselinestretch{1}\normalsize\small
\caption[]{The effective wavelength $\lambda_\mathrm{eff}$ (nm), number of stars and median precision of the photometry (mag) for the datasets and filters under consideration.
}
\label{filters}
\[
\begin{tabular}{llrrrr}
\hline
\noalign{\smallskip}
 Telescope, dataset, reference & Filter & $\lambda_\mathrm{eff}$ &  \multicolumn{3}{c}{Number of stars / Median precision} \\
\hline
\noalign{\smallskip}
           &        &                        & NGC\,288 & NGC\,362 & NGC\,6218 \\
\hline
\noalign{\smallskip}
{\it HST}/WFC3 \citep{nardiello2018}                                & $F275W$             & 285 & 3454 / 0.02      & 11757 / 0.02 & 5936 / 0.02  \\ 
{\it HST}/WFC3 \citep{nardiello2018}                                & $F336W$             & 340 & 5226 / 0.01      & 17105 / 0.02 & 8632 / 0.01  \\ 
1.54 m Danish Telescope, ESO, La Silla \citepalias{grundahl1999}    & Str\"omgren $u$     & 349 & 9566 / 0.03      & 5263 / 0.03  &             \\ 
Various \citepalias{stetson2019}                                    & $U$                 & 366 & 10144 / 0.02     &              & 12045 / 0.01   \\ 
1 m Las Campanas Swope telescope, MCPS \citep{zaritsky2002}         & $U$                 & 366 &                  & 2640 / 0.10  &             \\ 
1.54 m Danish Telescope, ESO, La Silla \citepalias{grundahl1999}    & Str\"omgren $v$     & 414 & 10936 / 0.02     & 6876 / 0.02  &             \\ 
{\it HST}/WFC3 \citep{nardiello2018}                                & $F438W$             & 438 & 4309 / 0.02      & 15348 / 0.02 & 10588 / 0.01  \\ 
{\it HST}/WFPC2 \citep{piotto2002}                                  & $B$ (from $F439W$)  & 452 &                  & 14325 / 0.04 & 4186 / 0.04   \\ 
Various \citepalias{stetson2019}                                    & $B$                 & 452 & 13910 / 0.01     &              & 12129 / 0.01   \\ 
1 m Las Campanas Swope telescope, MCPS \citep{zaritsky2002}         & $B$                 & 452 &                  & 6399 / 0.06  &              \\ 
2.2 m telescope, ESO, La Silla \citepalias{sollima2016}             & $B$                 & 452 & 9908 / 0.01      &              &              \\ 
4 m and 0.9 m telescopes, CTIO \citep{bolte1992}                    & $B$                 & 452 & fiducial / 0.03  &              &             \\ 
0.9 m telescope, KPNO \citepalias{hargis2004}                       & $B$                 & 452 &                  &              & 11292 / 0.04  \\ 
2.5 m du Pont telescope, Las Campanas \citepalias{zloczewski2012}   & $B$                 & 452 &                  &              & 7758 / 0.02   \\ 
1.54 m Danish Telescope, ESO, La Silla \citepalias{grundahl1999}    & Str\"omgren $b$     & 469 & 12372 / 0.02     &  7135 / 0.02 &             \\ 
4 m Blanco telescope, CTIO, DECam, DES \citep{des}                  & $g_\mathrm{DECam}$  & 481 & 2940 / 0.01      &              &              \\ 
{\it HST}/ACS \citep{simioni2018}                                   & $F475W$             & 484 &                  &              & 2776 / 0.01    \\ 
Pan-STARRS \citep{bernard2014,chambers2016}                         & $g_\mathrm{PS1}$    & 496 & 6902 / 0.02      &              & 10522 / 0.02   \\ 
SkyMapper Sky Survey DR3 \citep{onken2019}                          & $g_\mathrm{SMSS}$   & 514 & 2217 / 0.02      & 2194 / 0.02  & 2498 / 0.02  \\ 
{\it Gaia} DR2 \citep{gaiaevans}                                    & $G_\mathrm{BP}$     & 539 & 1713 / 0.03      & 1054 / 0.03  & 2847 / 0.02   \\ 
{\it Gaia} EDR3 \citep{riello2021}                                  & $G_\mathrm{BP}$     & 540 & 3923 / 0.03      & 4139 / 0.03  & 6231 / 0.03   \\ 
1.54 m Danish Telescope, ESO, La Silla \citepalias{grundahl1999}    & Str\"omgren $y$     & 550 & 12373 / 0.02     & 7023 / 0.03  &             \\ 
{\it HST}/WFPC2 \citep{piotto2002}                                  & $V$ (from $F555W$)  & 552 &                  & 14325 / 0.06 & 4186 / 0.03   \\ 
Various \citepalias{stetson2019}                                    & $V$                 & 552 & 13928 / 0.01     &              & 12129 / 0.01   \\ 
2.2 m telescope, ESO, La Silla \citepalias{sollima2016}             & $V$                 & 552 & 9908 / 0.01      &              &              \\ 
4 m and 0.9 m telescopes, CTIO \citep{bolte1992}                    & $V$                 & 552 & fiducial / 0.02  &              &              \\ 
2.2 m ESO/MPI telescope, La Silla, EFOSC2 \citepalias{bellazzini2001} & $V$               & 552 & 10777 / 0.02     & 7271 / 0.03  &              \\ 
1 m Las Campanas Swope telescope, MCPS \citep{zaritsky2002}           & $V$               & 552 &                  & 6399 / 0.06  &              \\ 
0.91 m DUTCH telescope, ESO, La Silla \citepalias{rosenberg2000}      & $V$               & 552 &                  & 1391 / 0.02  &             \\ 
0.9 m telescope, KPNO \citepalias{hargis2004}                       & $V$                 & 552 &                  &              & 14019 / 0.03  \\ 
2.5 m du Pont telescope, Las Campanas \citepalias{zloczewski2012}   & $V$                 & 552 &                  &              & 7758 / 0.01    \\ 
{\it HST}/ACS \citep{nardiello2018}                                 & $F606W$             & 599 & 13170 / 0.01     & 38951 / 0.01 & 20344 / 0.01   \\ 
SkyMapper Sky Survey DR3 \citep{onken2019}                          & $r_\mathrm{SMSS}$   & 615 & 2495 / 0.02      & 1917 / 0.02  & 3146 / 0.02 \\
Pan-STARRS \citep{bernard2014,chambers2016}                         & $r_\mathrm{PS1}$    & 621 & 7877 / 0.02      &              & 12195 / 0.01   \\ 
{\it Gaia} DR2 \citep{gaiaevans}                                    & $G$                 & 642 & 1713 / 0.01      & 1054 / 0.01  & 2847 / 0.01   \\ 
{\it Gaia} EDR3 \citep{riello2021}                                  & $G$                 & 642 & 3923 / 0.01      & 4139 / 0.01  & 6231 / 0.01   \\ 
4 m Blanco telescope, CTIO, DECam, DES \citep{des}                  & $r_\mathrm{DECam}$  & 644 & 3890 / 0.01      &              &             \\ 
Various \citepalias{stetson2019}                                    & $R$                 & 659 & 13726 / 0.01     &              & 12111 / 0.01   \\ 
Pan-STARRS \citep{bernard2014,chambers2016}                         & $i_\mathrm{PS1}$    & 752 & 7885 / 0.02      &              & 13030 / 0.02   \\ 
{\it Gaia} DR2 \citep{gaiaevans}                                    & $G_\mathrm{RP}$     & 767  & 1713 / 0.02     & 1054 / 0.02  & 2847 / 0.01   \\ 
{\it Gaia} EDR3 \citep{riello2021}                                  & $G_\mathrm{RP}$     & 774  & 3923 / 0.03     & 4139 / 0.03  & 6231 / 0.02   \\ 
SkyMapper Sky Survey DR3 \citep{onken2019}                          & $i_\mathrm{SMSS}$   & 776  & 2981 / 0.02     & 2570 / 0.03  & 4098 / 0.02 \\
4 m Blanco telescope, CTIO, DECam, DES \citep{des}                  & $i_\mathrm{DECam}$  & 784  & 4037 / 0.01     &              &             \\ 
{\it HST}/ACS \citep{nardiello2018}                                 & $F814W$             & 807  & 13170 / 0.01    & 38951 / 0.01 & 20344 / 0.01   \\ 
{\it HST}/ACS \citep{simioni2018}                                   & $F814W$             & 807  &                 &              & 2776 / 0.01   \\ 
Various \citepalias{stetson2019}                                    & $I$                 & 807  & 13898 / 0.01    &              & 12113 / 0.01  \\ 
2.2 m ESO/MPI telescope, La Silla, EFOSC2 \citepalias{bellazzini2001} & $I$               & 807  & 10777 / 0.03    & 7271 / 0.04  &             \\ 
1 m Las Campanas Swope telescope, MCPS \citep{zaritsky2002}         & $I$                 & 807  &                 & 5721 / 0.07  &             \\ 
0.91 m DUTCH telescope, ESO, La Silla \citepalias{rosenberg2000}    & $I$                 & 807  &                 & 1391 / 0.03  &             \\ 
0.9 m telescope, KPNO \citepalias{hargis2004}                       & $I$                 & 807  &                 &              & 13913 / 0.04  \\ 
Pan-STARRS \citep{bernard2014,chambers2016}                         & $z_\mathrm{PS1}$    & 867  & 6094 / 0.02     &              & 12578 / 0.02   \\ 
SkyMapper Sky Survey DR3 \citep{onken2019}                          & $z_\mathrm{SMSS}$   & 913  & 2375 / 0.03     & 1712 / 0.03  & 3435 / 0.02 \\
4 m Blanco telescope, CTIO, DECam, DES \citep{des}                  & $z_\mathrm{DECam}$  & 927  & 3373 / 0.02     &              &             \\ 
Pan-STARRS \citep{bernard2014,chambers2016}                         & $y_\mathrm{PS1}$    & 971  & 3706 / 0.03     &              & 6996 / 0.02   \\ 
3.6 m CFHT telescope \citep{davidge1997}                            & $J$                 & 1234 & fiducial / 0.05 &              &              \\ 
4 m Blanco telescope, CTIO, ISPI \citep{cohen2015}                  & 2MASS $J$           & 1235 &                 & 7412 / 0.02  &              \\ 
4 m Blanco telescope, CTIO, ISPI \citep{cohen2015}                  & 2MASS $K_s$         & 2176 &                 & 7412 / 0.03  &              \\ 
3.6 m CFHT telescope \citep{davidge1997}                            & $K$                 & 2193 & fiducial / 0.05 &              &              \\ 
{\it WISE}, unWISE \citep{unwise}                                   & $W1$                & 3317 & 794 / 0.03      & 764 / 0.01   & 1563 / 0.01   \\ 
{\it Spitzer}, IRAC, SAGE \citep{sage}                              & 3.6 $\mu$m          & 3524 &                 & 1559 / 0.09  &       \\ 
\hline
\end{tabular}
\]
\end{table*}

\begin{table}
\def\baselinestretch{1}\normalsize\normalsize
\caption[]{The fiducial sequences for NGC\,288, NGC\,362, and NGC\,6218 $G_\mathrm{RP}$ versus $G_\mathrm{BP}-G_\mathrm{RP}$ based on the data of
{\it Gaia} EDR3. The complete table is available online.
}
\label{fiducial}
\[
\begin{tabular}{cccccc}
\hline
\noalign{\smallskip}
\multicolumn{2}{c}{NGC\,288} & \multicolumn{2}{c}{NGC\,362} & \multicolumn{2}{c}{NGC\,6218} \\
\noalign{\smallskip}
$RP$ & $BP-RP$ & $RP$ & $BP-RP$ & $RP$ & $BP-RP$ \\
\hline
\noalign{\smallskip}
17.60 & -0.24 & 15.05 & 0.58 & 16.60 & 0.08 \\
17.20 & -0.20 & 14.90 & 0.72 & 16.05 & 0.12 \\
16.90 & -0.16 & 14.88 & 0.76 & 15.70 & 0.16 \\
16.60 & -0.12 & 14.86 & 0.80 & 15.40 & 0.20 \\
16.20 & -0.06 & 14.82 & 0.84 & 15.20 & 0.24 \\
\ldots & \ldots & \ldots & \ldots & \ldots & \ldots \\
\hline
\end{tabular}
\]
\end{table}

To be fitted by a theoretical isochrone, data should be presented by a fiducial sequence, i.e. a colour--magnitude relation for single stars. 
Such a sequence is calculated as a locus of the number density maxima in some colour--magnitude bins. Some details and examples are given in 
section 3 of \citetalias{ngc6205}.

The stellar populations are segregated at the HB and AGB in almost all CMDs with the UV and optical filters, as well as
at the RGB in the following CMDs\footnote{We find no CMD for these GCs with the populations segregated at the SGB, TO, or MS.}: 
$F275W-F336W$, $F336W-F438W$, $U-B$, $R-I$, Str\"omgren $u-v$ for all the GCs and $F438W-F606W$ for NGC\,288.
In such a case, each population is presented by its own fiducial sequence, which, in turn, is fitted by its own primordial or helium-enriched isochrone
(the DSEP helium-enriched isochrone with $Y=0.275$ is interpolated in each CMD from the isochrones with $Y=0.2475$ and 0.33
\footnote{Since the initial isochrones are close to each other in any CMD and since they are presented by the same evolutionary points,
such interpolation in colour--magnitude planes is robust, with uncertainties less than 0.01 mag.}).
For the remaining cases, we consider and fit an unresolved mix of the populations. 
Since two populations, with $Y\approx0.248$ and $Y\approx0.275$, are well presented in these GCs \citep{milone2017}, it is reasonable to 
adopt the average value $Y\approx0.26$ for their mix.
This mix is presented by a fiducial sequence, which, in turn, is fitted by an isochrone for $Y=0.26$ interpolated in each CMD between the primordial and 
helium-enriched isochrones of the same model, distance, reddening, and age\footnote{The helium-enriched population may be a few hundred Myr younger, 
but this age difference is negligible w.r.t. precision of the derived ages, indicated in Sect.~\ref{results}.}
(see Sect.~\ref{iso}).

For our clusters, some important CMD domains have a small number of stars: e.g. red HBs of NGC\,288 and NGC\,6218.
In such cases, a fiducial point is defined by a few or even only one star, if the photometry of such star/stars is rather precise, i.e.
if a colour and a magnitude of such a fiducial point can be defined within $\pm0.04$ mag.
Since we use a lot of fiducial points to derive distance, age, and reddening, such an uncertainty of a fiducial point is negligible w.r.t.
total uncertainty (see the balance of uncertainties in appendix A of \citetalias{ngc6205}).

As an example, the fiducial sequences for the {\it Gaia} EDR3 datasets are presented in Table~\ref{fiducial}.
All other fiducial sequences can be provided on request.

Similar to \citetalias{ngc6205}, we analyse variations of the fiducial sequence colour over the cluster fields for all the CMDs, which are based on 
the largest datasets of {\it HST} ACS/WFC3, Pan-STARRS, \citetalias{grundahl1999} and \citetalias{stetson2019}.
We calculate partial fiducial sequences, i.e. the ones for different parts of the field, along the RGB, SGB, TO and a brighter part of the MS by use 
of a moving window with about 2000 stars. As for NGC\,6205 in \citetalias{ngc6205}, we find significant variations of the fiducial sequence colour 
only for some datasets and only in most crowded central parts of GCs, up to 2 arcmin. In such cases, stars in the central parts are removed.
Elsewhere, we find rather small fiducial sequence colour variations at a level of $\Delta(B-V)<0.02$ mag.
Such a level agrees with the estimates of \citet{anderson2008} as well as with a rather small mean differential reddening $\overline{\delta E(B-V)}$ 
in the fields of NGC\,362 and NGC\,6218, obtained by \citetalias{bonatto2013} and presented in Table~\ref{properties}.
Given such a small effect for the largest datasets, we suggest a similar effect for the others.
As shown in \citetalias{ngc6205}, since we consider the entire cluster fields, this effect is averaged. Therefore, this makes a negligible 
contribution to the total uncertainties and, hence, we do not take it into account further.

\subsection{Cleaning the datasets}
\label{cleaning}

\begin{figure}
\includegraphics{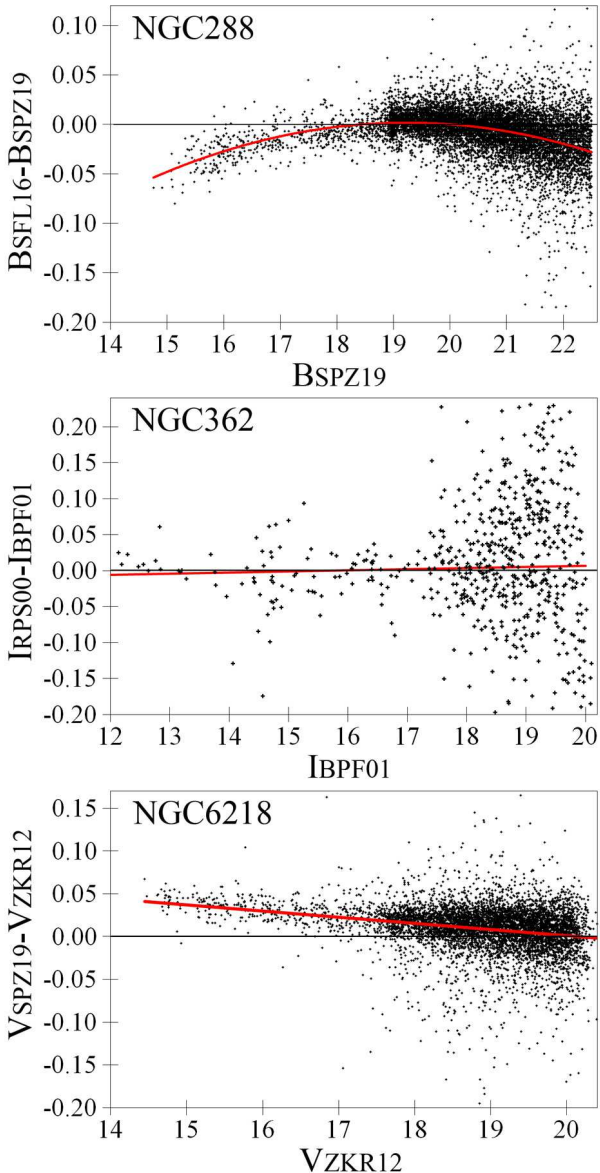}
\caption{Some examples of dataset magnitude differences as functions of magnitude. The approximating functions from 
Table~\ref{bvitable} are shown by the red curves.
}
\label{bvi}
\end{figure}

Cleaning the original datasets from poor photometry for our GCs is similar to what was done for NGC\,6205 in \citetalias{ngc6205}. 

We select stars with a photometric error of less than 0.1~mag.
A higher limit of $<0.15$ mag is applied to MCPS and SAGE due to their less precise photometry (see Table~\ref{filters}).

For the datasets of \citetalias{stetson2019}, \citetalias{grundahl1999}, \citetalias{sollima2016}, \citet{piotto2002}, \citetalias{rosenberg2000} 
and some others, we select stars with \verb"DAOPHOT" \verb"sharp" parameters $|{\tt sharp}|<0.3$ and $\chi<3$.
For the HST WFC3 and ACS photometry we select stars with $|{\tt sharp}|<0.15$, membership probability $>0.9$ or $-1$, and quality fit $>0.9$.
For DES we select stars with extended source flags $>0.6$ in all the DES filters.
For Pan-STARRS we select stars with the parameter `Maximum point-spread function weighted fraction of pixels totally unmasked from 
filter detections' $>0.9$ for all the filters under consideration.
For the {\it Gaia} DR2 dataset, we select stars with a precise photometry as those with available data in all three {\it Gaia} bands and 
with an acceptable parameter \verb"phot_bp_rp_excess_factor"$>1.3+0.06$\verb"bp_rp"$^2$, as suggested by \citet{gaiaevans}.

For each cluster we have several datasets with photometry in the $BVI$ filters. We cross-identify these datasets for a direct comparison.
Their magnitude differences are presented as some functions of magnitude in Table~\ref{bvitable}. 
Some examples of these functions are shown in Fig.~\ref{bvi} by the red curves.
Such magnitude differences at a level of few hundredths of a magnitude are common (see a detailed analysis by \citealt{anderson2008}). 
As we do not know which dataset has the correct photometric system, we have to put the datasets on an average system.
To do this, we use the functions from Table~\ref{bvitable} in order to correct the magnitudes from these datasets keeping the same average magnitudes of 
their common stars. All the corrections are within a few hundredths of a magnitude. 
Since the datasets of \citet{bolte1992} and \citet{piotto2002} cannot be cross-identified with the others, no correction is applied to them.
We do not use them for creating the average photometric system, but we use them as is in the isochrone fitting.

\begin{table*}
\def\baselinestretch{1}\normalsize\normalsize
\caption[]{Pairs of datasets with $B$, $V$, and $I$ filters: their differences as functions of magnitude.
}
\label{bvitable}
\[
\begin{tabular}{lccc}
\hline
\noalign{\smallskip}
Dataset pair & $\Delta B$ & $\Delta V$ & $\Delta I$ \\
\hline
\noalign{\smallskip}
             &            & NGC\,288   &            \\
\citetalias{sollima2016}-\citetalias{stetson2019}    & $-0.0028B^2+0.1067B-1.0245$ & $+0.0010V-0.0194$ &    \\
\citetalias{bellazzini2001}-\citetalias{stetson2019} &                             & $-0.0050V+0.0950$ & $+0.0008I-0.03$ \\
\citetalias{bellazzini2001}-\citetalias{sollima2016} &                             & $-0.0060V+0.1144$ & \\
\hline
\noalign{\smallskip}
             &            & NGC\,362   &            \\
MCPS-\citetalias{rosenberg2000}                        &             & $+0.0024V-0.0649$   &  $-0.0002I-0.0035$ \\
\citetalias{rosenberg2000}-\citetalias{bellazzini2001} &             & $+0.0011V-0.0134$   &  $+0.0015I-0.0236$ \\
MCPS-\citetalias{bellazzini2001}                       &             & $+0.0035V-0.0783$   &  $+0.0013I-0.0271$ \\
\hline
\noalign{\smallskip}
             &            & NGC\,6218   &            \\
\citetalias{stetson2019}-\citetalias{zloczewski2012}  & $-0.0047B+0.0628$       & $-0.0069V+0.1396$   &   \\
\citetalias{hargis2004}-\citetalias{stetson2019}      & $-0.0010B+0.0218$       & $+0.0006V-0.0178$   & $+0.0009I-0.0191$ \\
\citetalias{hargis2004}-\citetalias{zloczewski2012}   & $-0.0057B+0.0846$       & $-0.0063V+0.1218$   & \\
\hline
\end{tabular}
\]
\end{table*}

\begin{figure}
\includegraphics{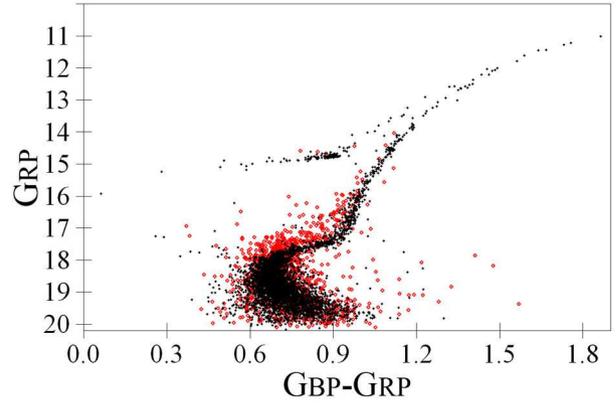}
\caption{$G_\mathrm{BP}-G_\mathrm{RP}$ versus $G_\mathrm{RP}$ CMD for NGC\,362 stars with acceptable or unacceptable BP/RP excess factor, which are
shown by black and red symbols, respectively, after applying the remaining cleaning of the sample.
}
\label{ngc362gaiaclean1}
\end{figure}

\begin{figure}
\includegraphics{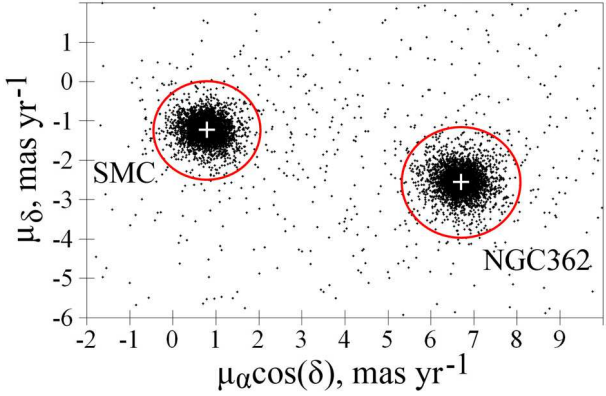}
\caption{The distribution of stars, selected within 18.7 arcmin of the NGC\,362 centre, on the proper motion components (mas\,yr$^{-1}$), 
after the remaining cleaning of the sample.
The weighted mean PM and the selection area are shown by the white cross and red circle for the SMC (left) and NGC\,362 (right).
}
\label{ngc362pm}
\end{figure}

\begin{figure}
\includegraphics{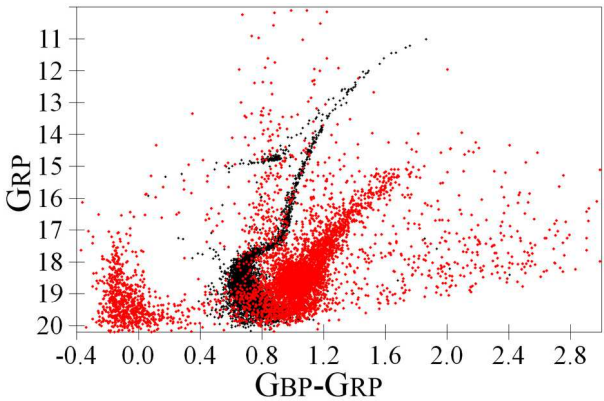}
\caption{$G_\mathrm{BP}-G_\mathrm{RP}$ versus $G_\mathrm{RP}$ CMD for NGC\,362 stars selected by their proper motion (black symbols) versus
the SMC and other foreground and background stars (red symbols) after the remaining cleaning of the sample.
}
\label{ngc362gaiaclean0}
\end{figure}

\subsection{Cleaning the {\it Gaia} EDR3 datasets}
\label{edr3}

Comparing {\it Gaia} DR2 with EDR3, we note that the latter, being a magnitude deeper, provides many more cluster stars, as is evident from 
Table~\ref{filters}.
Also, {\it Gaia} EDR3 provides more precise parallaxes and PMs for these stars.
This allows us to do a special cleaning of the {\it Gaia} EDR3 datasets.

We remove {\it Gaia} EDR3 stars with \verb"duplicated_source"$=1$ (\verb"Dup=1"), 
\verb"astrometric_excess_noise"$>1$ ($\epsilon i>1$), or renormalised unit weight error exceeding $1.2$ (\verb"RUWE"$>1.2$).
We remove foreground and background stars as those with measured parallax $\varpi>1/R+3\sigma_{\varpi}$ or $\varpi<1/R-3\sigma_{\varpi}$, 
where $R$ is the distance to the cluster derived by us (it is upgraded iteratively starting from the \citealt{harris} values in 
Table~\ref{properties}), and $\sigma_{\varpi}$ is the stated parallax uncertainty.
Also, we remove few stars without PMs.
We select {\it Gaia} EDR3 stars with precise photometry as those with available data in all three {\it Gaia} bands and with an acceptable 
parameter \verb"phot_bp_rp_excess_factor_corrected" (i.e. \verb"E(BP/RP)Corr") between $-0.14$ and $0.14$ \citep{riello2021}.
Fig.~\ref{ngc362gaiaclean1} shows the {\it Gaia} EDR3 stars of NGC\,362 with acceptable and unacceptable \verb"phot_bp_rp_excess_factor_corrected" 
as black and red symbols, respectively, after applying the remaining cleaning of the sample. 
It is seen that the stars with unacceptable \verb"phot_bp_rp_excess_factor_corrected" deviate systematically.

This cleaning removes almost all stars in a few central arcminutes of the cluster fields. 
The remaining stars in the centres do not show any systematics in the CMDs.

We select likely cluster members and derive systemic PMs of the clusters based on the positions and PM components $\mu_{\alpha}\cos(\delta)$ 
and $\mu_{\delta}$ of the stars. This procedure includes the following steps:
\begin{enumerate}
\item The initial cluster centre coordinates are adopted from \citet{goldsbury2010}.
\item The initial cluster systemic PM components $\overline{\mu_{\alpha}\cos(\delta)}$ and $\overline{\mu_{\delta}}$ are adopted from 
\citet[][hereafter VB21]{vasiliev2021}.
\item The initial {\it Gaia} EDR3 sample is selected within 45, 45, and 57 arcmin from the centres of NGC\,288, NGC\,362, and NGC\,6218, 
respectively.
\item We calculate the deviations of the individual PMs from the systemic PM as 
$\sqrt{(\mu_{\alpha}\cos(\delta)-\overline{\mu_{\alpha}\cos(\delta)})^2+(\mu_{\delta}-\overline{\mu_{\delta}})^2}$.
\item Initially, we remove stars with 
$\sqrt{(\mu_{\alpha}\cos(\delta)-\overline{\mu_{\alpha}\cos(\delta)})^2+(\mu_{\delta}-\overline{\mu_{\delta}})^2}>2$ mas\,yr$^{-1}$, 
since the PM dispersion and stated {\it Gaia} EDR3 PM uncertainties specify a $3\sigma$ limit, which rejects these stars as being unlikely cluster members.
\item We find the cluster truncation radius at a sharp drop of the cluster radial star number density profile, and select cluster members 
as stars within the truncation radius. 
\item We calculate the standard deviations $\sigma_{\mu_{\alpha}\cos(\delta)}$ and $\sigma_{\mu_{\delta}}$ of the PM components of the members.
\item We cut the sample at $3\sigma$, i.e. only select stars with 
$\sqrt{(\mu_{\alpha}\cos(\delta)-\overline{\mu_{\alpha}\cos(\delta)})^2+(\mu_{\delta}-\overline{\mu_{\delta}})^2}<3\sqrt{\sigma_{\mu_{\alpha}\cos(\delta)}^2+\sigma_{\mu_{\delta}}^2}$
as cluster members.
\item We recalculate the mean coordinates of the cluster centre.
\item We recalculate the systemic PM components as the weighted means of the individual PM components of the cluster members.
\item The steps (iv) and (vi)--(x) are repeated iteratively.
\end{enumerate}
This procedure converges after several iterations. The final empirical standard deviations 
$\sigma_{\mu_{\alpha}\cos(\delta)}$ and $\sigma_{\mu_{\delta}}$ appear reasonable, being slightly higher than the mean stated {\it Gaia} EDR3 
PM uncertainties: 0.36 versus 0.29, 0.34 versus 0.25, and 0.35 versus 0.27 mas\,yr$^{-1}$ for NGC\,288, NGC\,362, and NGC\,6218, respectively
(averaged for the PM components).

The final truncation radii are 13.7, 18.7, and 14.9 arcmin for NGC\,288, NGC\,362, and NGC\,6218, respectively.
They are larger than those provided in Table~\ref{properties} -- 7.5, 7.5, and 9.5 arcmin, respectively.
This increase of the radii adds 9, 19, and 7 per cent of members for NGC\,288, NGC\,362, and NGC\,6218, respectively.
However, they cannot be selected as cluster members without knowing their precise PMs.
Anyway, these additional members do not change our results.

This cleaning is especially important for NGC\,362 whose field is contaminated by SMC stars with different PMs.
Fig.~\ref{ngc362pm} shows the final distribution of stars, selected within 18.7 arcmin of the NGC\,362 centre, by the proper motion components.
A clear separation of the SMC and NGC\,362 is evident. The $3\sigma$ cut, shown by the red circles, seems to be reasonable. 
The CMD of these stars is shown in Fig.~\ref{ngc362gaiaclean0}.

Since the stated {\it Gaia} EDR3 PM uncertainty strongly increases with magnitude, faint cluster members 
contribute negligibly to the weighted mean PMs.

\begin{table}
\def\baselinestretch{1}\normalsize\normalsize
\caption[]{The cluster systemic PMs (mas\,yr$^{-1}$).
}
\label{systemic}
\[
\begin{tabular}{llcc}
\hline
\noalign{\smallskip}
Cluster & Source & $\mu_{\alpha}\cos(\delta)$ & $\mu_{\delta}$ \\
\hline
\noalign{\smallskip}
          & This study           & $4.147\pm0.005$ & $-5.704\pm0.006$ \\
NGC\,288  & \citetalias{vasiliev2021} & $4.164\pm0.024$ & $-5.705\pm0.025$ \\
          & \citet{vitral2021}   & $4.154\pm0.005$ & $-5.700\pm0.005$ \\
\hline
\noalign{\smallskip}
          & This study           & $6.696\pm0.005$ & $-2.543\pm0.005$ \\
NGC\,362  & \citetalias{vasiliev2021} & $6.694\pm0.024$ & $-2.536\pm0.024$ \\
          & \citet{vitral2021}   & $6.680\pm0.011$ & $-2.545\pm0.011$ \\
\hline
\noalign{\smallskip}
          & This study           & $-0.204\pm0.005$ & $-6.809\pm0.004$ \\
NGC\,6218 & \citetalias{vasiliev2021} & $-0.191\pm0.024$ & $-6.801\pm0.024$ \\
          & \citet{vitral2021}   & $-0.213\pm0.007$ & $-6.811\pm0.007$  \\
\hline
\end{tabular}
\]
\end{table}

Our final weighted mean PMs are presented in Table~\ref{systemic}. They are compared with the estimates from \citetalias{vasiliev2021}
and \citet{vitral2021}, which are also based on {\it Gaia} EDR3.
Despite different approaches used, these estimates are consistent within $\pm0.01$ mas\,yr$^{-1}$.
However, the {\it Gaia} EDR3 PM systematic errors may be higher, equally affecting all these estimates.
The random uncertainty is indicated in Table~\ref{systemic} for our and \citet{vitral2021} estimates, while the total (random plus systematic)
uncertainty -- for those of \citetalias{vasiliev2021}.
\citetalias{vasiliev2021} find that these total uncertainty values should be considered as `the irreducible systematic floor' on the accuracy of the 
{\it Gaia} EDR3 PMs `for any compact stellar system', while 0.011 mas is a respective limit for its parallaxes.
Hence, we adopt these values as the final uncertainties of our PMs, as well as of median parallaxes from {\it Gaia} EDR3.
The latter are calculated by us with the parallax zero-point correction from \citet{gaia2021c} applied.
We present these parallaxes in Table~\ref{properties} and compare them with our results in Sect.~\ref{agedist}.

Note that GC systemic PM estimates based on {\it Gaia} EDR3 are most accurate to date. This is seen in their comparison with previous estimates of PMs, 
e.g. those recently obtained for NGC\,362 by \citet{libralato2018} using {\it HST} data: $\mu_{\alpha}\cos(\delta)=6.703\pm0.278$, 
$\mu_{\delta}=-2.407\pm0.135$ mas\,yr$^{-1}$.

The final lists of the {\it Gaia} EDR3 cluster members are presented in Table~\ref{gaiaedr3}.

\begin{table*}
\def\baselinestretch{1}\normalsize\normalsize
\caption[]{The list of the {\it Gaia} EDR3 members of NGC\,288, NGC\,362, and NGC\,6218. The complete table is available online.
}
\label{gaiaedr3}
\[
\begin{tabular}{ccc}
\hline
\noalign{\smallskip}
NGC\,288 & NGC\,362 & NGC\,6218 \\
\hline
\noalign{\smallskip}
Gaia EDR3 2342707263271857664 & Gaia EDR3 4690797804192972288 & Gaia EDR3 4379051313863587968 \\
Gaia EDR3 2342707267566820864 & Gaia EDR3 4690797941632010624 & Gaia EDR3 4379054165721868544 \\
Gaia EDR3 2342712863909223168 & Gaia EDR3 4690798010351505792 & Gaia EDR3 4379054268801095296 \\
Gaia EDR3 2342713070067644032 & Gaia EDR3 4690809383424866560 & Gaia EDR3 4379054513620043648 \\
Gaia EDR3 2342713413665033600 & Gaia EDR3 4690809795741849344 & Gaia EDR3 4379054578038751232 \\
\ldots & \ldots & \ldots \\
\hline
\end{tabular}
\]
\end{table*}

By use of the same approach, we select 5052 members of the SMC, which represent a background of NGC\,362. They are presented in Fig.~\ref{ngc362pm}.
We calculate their systemic PM:
$$\mu_{\alpha}\cos(\delta)=0.786\pm0.004\pm0.026,\, \mu_{\delta}=-1.222\pm0.004\pm0.026$$
mas\,yr$^{-1}$, where the former and latter errors are the random and systematic uncertainties, respectively. 
Note that due to the rotation of the SMC, this is the PM of a part of the SMC just behind NGC\,362.
Other parts may have a different PM.

To clean the SMSS datasets, we cross-identify them with the lists of the {\it Gaia} EDR3 members of these GCs and consider only common stars.
This is an efficient and accurate segregation of the cluster members and contaminants, 
since both the SMSS and the {\it Gaia} EDR3 datasets are rather complete, deep and have many common stars.

\section{Theoretical models and isochrones}
\label{iso}

In order to fit the CMDs of NGC\,288, NGC\,362, and NGC\,6218, we use the following theoretical models of stellar evolution and related 
pairs of $\alpha$--enhanced isochrones:
\begin{enumerate}
\item BaSTI-IAC \citep{newbasti,pietrinferni2021} with [Fe/H]$=-1.30$, [$\alpha$/Fe]$=+0.40$, initial solar $Z_{\sun}=0.0172$ and $Y_{\sun}=0.2695$, 
overshooting, diffusion, mass loss efficiency $\eta=0.3$, where $\eta$ is the free parameter in Reimers law \citep{reimers}.
We adopt $Z=0.001572$, $Y=0.249$ for primordial and $Z=0.001570$, $Y=0.275$ for helium-enriched population.
The interpolated isochrone for $Y=0.26$ is between them. It is used to describe the observed mix of the populations.
We also use the BaSTI-IAC extended set of zero-age horizontal branch (ZAHB) models with different values of the total mass but the same mass 
for the helium core and the same envelope chemical stratification. 
This set takes into account the assumption that stars with the same mass during the MS can lose different amount of mass during the RGB and, hence, 
differ in their colours and magnitudes during the HB.
The BaSTI-IAC extended set of ZAHB models seems to be the most appropriate theoretical representation of the HB to date.
\item DSEP \citep{dotter2008} with [Fe/H]$=-1.32$, [$\alpha$/Fe]$=+0.40$, solar $Z_{\sun}=0.0189$ and no mass loss.
We use the isochrones with $Z=0.001564$, $Y=0.2475$ and $Z=0.001388$, $Y=0.33$. DSEP provides no $\alpha$--enhanced isochrone for $0.2475<Y<0.33$. 
Therefore, we use the isochrone with $Y=0.2475$ to describe the primordial population, while interpolating in each CMD the isochrone with $Y=0.275$ for 
the helium-enriched population and the isochrone with $Y=0.26$ for a mix of the populations.
Naturally, both the interpolated isochrones are much closer to that of $Y=0.2475$ than that of $Y=0.33$.
DSEP does not provide the HB and AGB.
\end{enumerate}

DSEP and BaSTI-IAC are the only models providing $\alpha$--enhanced isochrones for almost all the filters under consideration. 
These models have demonstrated the most extreme results among all the models used in \citetalias{ngc5904} and \citetalias{ngc6205}: 
the lowest ages and the highest reddenings by DSEP, while the highest ages and the lowest reddenings by BaSTI-IAC.
This can be explained if one compares the models in a diagram of effective temperature $T_\mathrm{eff}$ versus luminosity, i.e. before applying 
the colour--$T_\mathrm{eff}$ relations and bolometric corrections. 
Such a comparison presented by \citet{pietrinferni2021} or in figure 13 of \citetalias{ngc6205} shows that the DSEP RGB and TO are about 100 K hotter, 
while the DSEP SGB is slightly shorter than those of BaSTI-IAC.
This is due to the differences in the physics inputs, most notably, in boundary conditions, solar metal mixture, and calibration of the solar
standard model \citep{newbasti, pietrinferni2021}.
Note that DSEP and BaSTI-IAC show similar SGB luminosities and, consequently, provide nearly the same distances.
Thus, DSEP and BaSTI-IAC are the most interesting models for our study.
The diversity of all possible models should be covered by these two extreme models.

However, we have checked whether various solar-scaled models can substitute $\alpha$--enhanced ones as suggested by the rule of \citet{salaris1993}: 
`$\alpha$--enhanced isochrones are very well mimicked by the standard scaled solar isochrones of metallicity $Z=Z_0(0.638f_\mathrm{\alpha}+0.362)$,
where $f_\mathrm{\alpha}$ is the chosen average enhancement factor of the $\alpha$-elements and $Z_0$ is the initial (nonenhanced) metallicity'.
For our GCs this means that solar-scaled isochrones with [Fe/H]$\approx-1$ and $\alpha$--enhanced isochrones with [Fe/H]$=-1.3$ would provide 
the same age, distance, and reddening.
\citet{pietrinferni2021} have found that the BaSTI-IAC isochrones for short wavelength filters significantly deviate from this rule.
According to this, our testing of various models has not revealed a model following this rule in \emph{all} the CMDs under consideration. 
Thus, we decide to use only DSEP and BaSTI-IAC $\alpha$--enhanced isochrones.

For {\it Gaia} DR2 and EDR3 we use isochrones based on the response curves of $G$, $G_\mathrm{BP}$, and $G_\mathrm{RP}$ from \citet{gaiaevans} 
and \citet{riello2021}, respectively.

We consider the isochrones for a grid of some reasonable distances with a step of 0.1 kpc, reddenings with a step of 0.001 mag, 
and ages over 8 Gyr with a step of 0.5 Gyr.
Similar to \citetalias{ngc6205}, to derive the most probable reddening, distance, and age, we select an isochrone with a minimal total offset 
between the isochrone points and the fiducial points in the same magnitude range of the CMD.

\begin{table}
\def\baselinestretch{1}\normalsize\normalsize
\caption{The results of the isochrone fitting for various models and some key colours for NGC\,288. The complete table is available online.
The {\it Gaia} DR2 results are shown for DSEP, while those from EDR3 -- for BaSTI-IAC.
}
\label{ngc288results}
\[
\begin{tabular}{lcc}
\hline
                                      & DSEP &  BaSTI-IAC  \\
\hline
$E(G_\mathrm{BP}-G_\mathrm{RP})$ EDR3/DR2 & $0.061\pm0.03$ & $0.027\pm0.03$ \\
age, Gyr                                  &   12.0         & 15.0     \\
distance, kpc                             &   9.0          & 8.8      \\
\noalign{\smallskip}
$E(B-V)$ \citetalias{stetson2019}         & $0.012\pm0.02$ & $-0.012\pm0.03$ \\
age, Gyr                                  & 12.5           & 15.5           \\
distance, kpc                             & 9.1            & 8.9             \\
\noalign{\smallskip}
$E(b-y)$                                  & $0.022\pm0.02$ & $0.005\pm0.02$ \\
age, Gyr                                  &  12.5          &  14.5     \\
distance, kpc                             &   8.6          &  8.9       \\
\ldots & \ldots & \ldots \\
\hline
\end{tabular}
\]
\end{table}

\begin{table}
\def\baselinestretch{1}\normalsize\normalsize
\caption{The same as Table~\ref{ngc288results} but for NGC\,362. The complete table is available online.
}
\label{ngc362results}
\[
\begin{tabular}{lcc}
\hline
                                      & DSEP & BaSTI-IAC \\
\hline
$E(G_\mathrm{BP}-G_\mathrm{RP})$ EDR3/DR2 & $0.085\pm0.03$ & $0.051\pm0.02$ \\
age, Gyr                                  &   10.5         & 11.5     \\
distance, kpc                             &   8.6          & 8.8  \\
\noalign{\smallskip}
$E(B-V)$ MCPS                             & $0.049\pm0.02$ & $0.027\pm0.02$ \\
age, Gyr                                  & 10.5           & 11.5 \\
distance, kpc                             & 8.8            & 9.0 \\
\noalign{\smallskip}
$E(b-y)$                                  & $0.040\pm0.03$ & $0.026\pm0.03$ \\
age, Gyr                                  &  10.0          &  10.5     \\
distance, kpc                             &   8.6          &  8.9  \\
\ldots & \ldots & \ldots \\
\hline
\end{tabular}
\]
\end{table}

\begin{table}
\def\baselinestretch{1}\normalsize\normalsize
\caption{The same as Table~\ref{ngc288results} but for NGC\,6218. The complete table is available online.
}
\label{ngc6218results}
\[
\begin{tabular}{lcc}
\hline
                                      & DSEP     & BaSTI-IAC \\
\hline
$E(G_\mathrm{BP}-G_\mathrm{RP})$ EDR3/DR2 & $0.295\pm0.03$ & $0.268\pm0.02$ \\
age, Gyr                                  &   13.0         & 15.5     \\
distance, kpc                             &   4.8          & 4.8  \\
\noalign{\smallskip}
$E(B-V)$ \citetalias{stetson2019}         & $0.202\pm0.02$ & $0.183\pm0.02$ \\
age, Gyr                                  & 13.0           & 15.5           \\
distance, kpc                             & 4.8            & 4.8             \\
\noalign{\smallskip}
$E(B-V)$ \citetalias{zloczewski2012}      & $0.204\pm0.02$ & $0.190\pm0.02$ \\
age, Gyr                                  & 12.5           & 14.0           \\
distance, kpc                             & 4.9            & 4.9             \\
\ldots & \ldots & \ldots \\
\hline
\end{tabular}
\]
\end{table}

\begin{figure}
\includegraphics{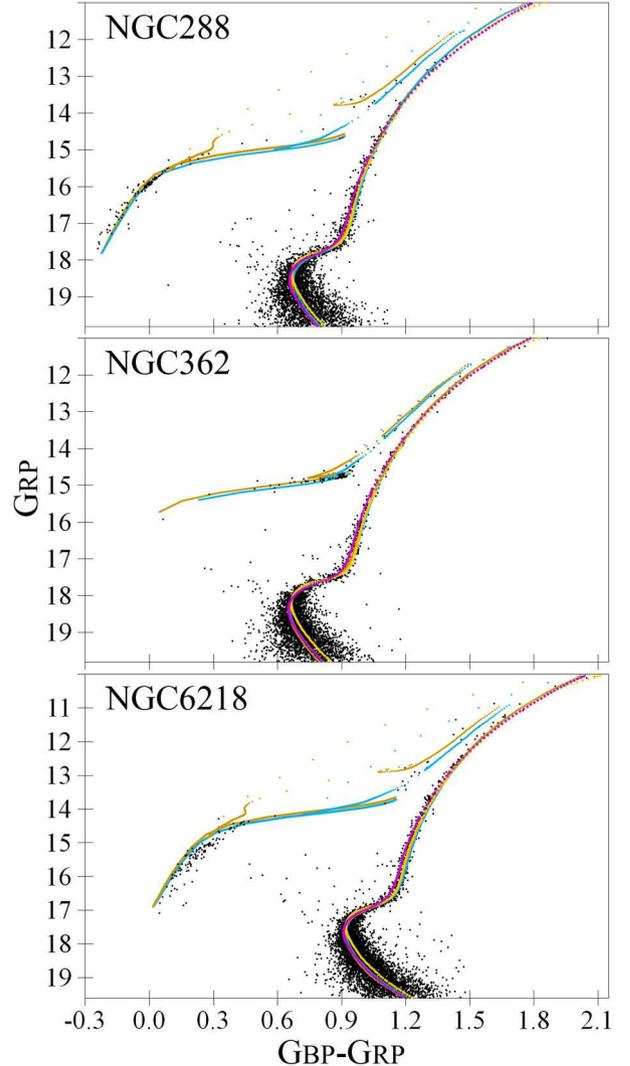}
\caption{{\it Gaia} EDR3 $G_\mathrm{BP}-G_\mathrm{RP}$ versus $G_\mathrm{RP}$ CMD of NGC\,288, NGC\,362, and NGC\,6218.
The isochrones from BaSTI-IAC for $Y=0.249$ (blue) and 0.275 (brown) and from DSEP for $Y=0.2475$ (yellow) and 0.33 (purple)
are calculated with the best-fitting parameters from Tables~\ref{ngc288results}, \ref{ngc362results}, and \ref{ngc6218results}, respectively. 
The DSEP isochrones are calculated for {\it Gaia} DR2, but they are shown here, since we find no significant systematic difference between the 
DR2 and EDR3 datasets.
}
\label{cmd01}
\end{figure}

\begin{figure}
\includegraphics{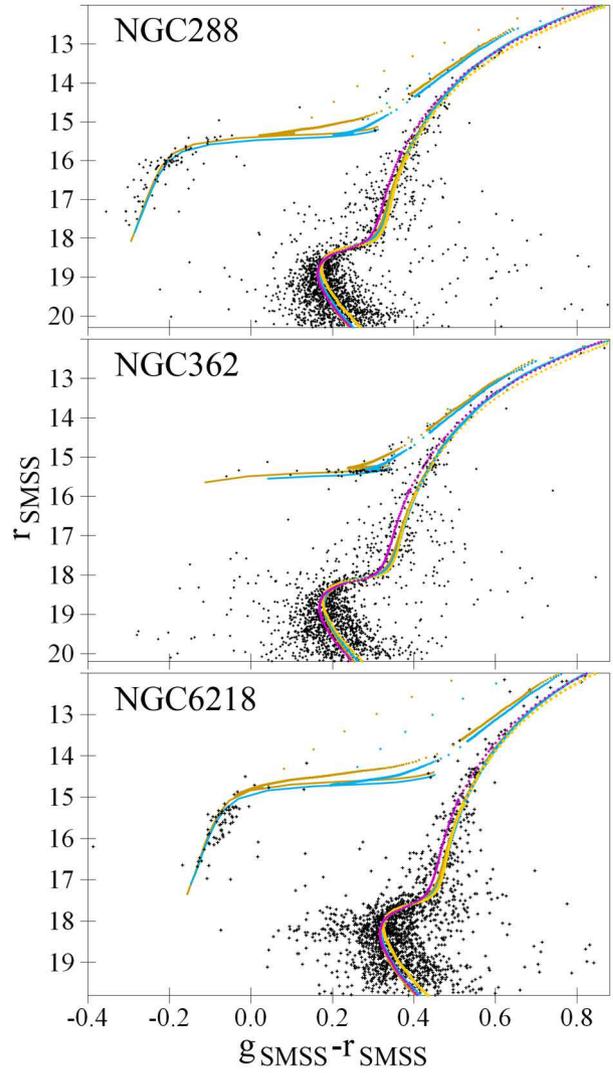}
\caption{The same as Fig.~\ref{cmd01} but for $g_\mathrm{SMSS}-r_\mathrm{SMSS}$ versus $r_\mathrm{SMSS}$ CMD of NGC\,288, NGC\,362, and NGC\,6218 
based on the data from SMSS.
The isochrones are calculated with the best-fitting parameters from Tables~\ref{ngc288results}, \ref{ngc362results}, and \ref{ngc6218results}, respectively.
}
\label{cmd02}
\end{figure}

\begin{figure}
\includegraphics{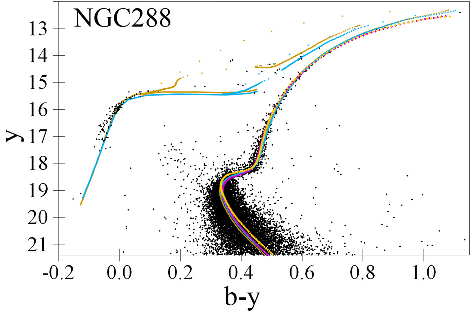}
\caption{The same as Fig.~\ref{cmd01} but for $b-y$ versus $y$ CMD of NGC\,288 based on the data from \citetalias{grundahl1999}.
The isochrones are calculated with the best-fitting parameters from Table~\ref{ngc288results}.
}
\label{cmd03}
\end{figure}

\begin{figure}
\includegraphics{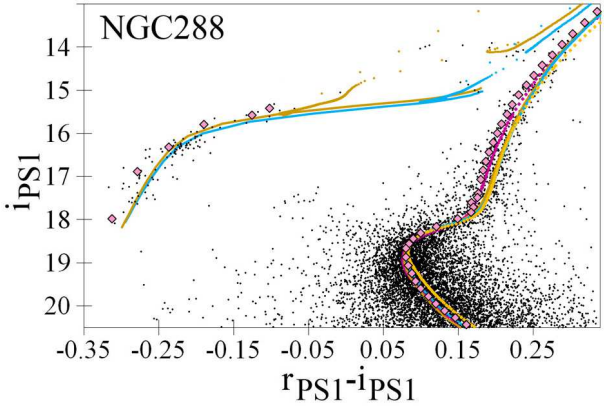}
\caption{The same as Fig.~\ref{cmd01} but for Pan-STARRS $r_\mathrm{PS1}-i_\mathrm{PS1}$ versus $i_\mathrm{PS1}$ CMD of NGC\,288 with the fiducial 
sequence of \citet{bernard2014} shown as the pink diamonds.
The isochrones are calculated with the best-fitting parameters from Table~\ref{ngc288results}.
}
\label{cmd04}
\end{figure}

\begin{figure}
\includegraphics{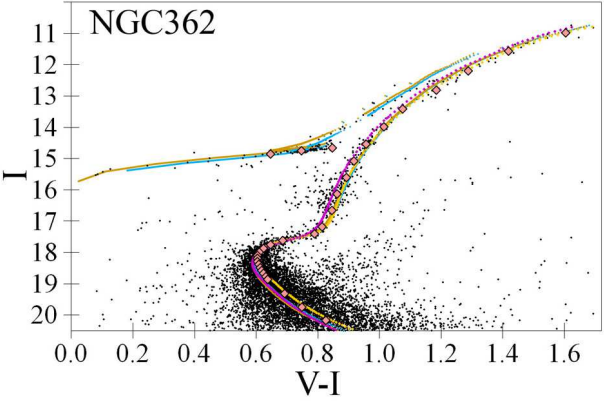}
\caption{The same as Fig.~\ref{cmd01} but for $V-I$ versus $I$ CMD of NGC\,362 based on the data from \citetalias{bellazzini2001} with their 
fiducial sequence shown as the pink diamonds.
The isochrones are calculated with the best-fitting parameters from Table~\ref{ngc362results}.
}
\label{cmd05}
\end{figure}

\begin{figure}
\includegraphics{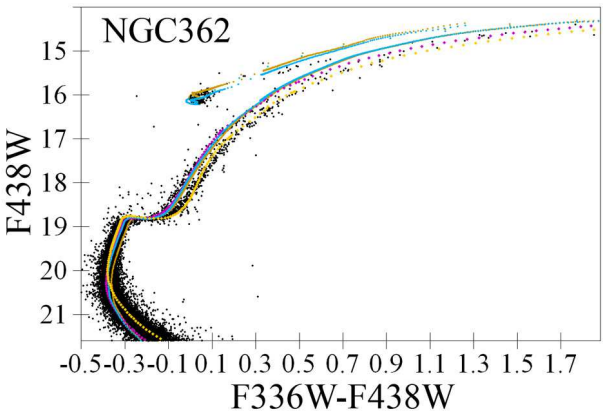}
\caption{The same as Fig.~\ref{cmd01} but for $F336W-F438W$ versus $F438W$ CMD of NGC\,362 based on the {\it HST}/WFC3 data.
The isochrones are calculated with the best-fitting parameters from Table~\ref{ngc362results}.
}
\label{cmd06}
\end{figure}

\begin{figure}
\includegraphics{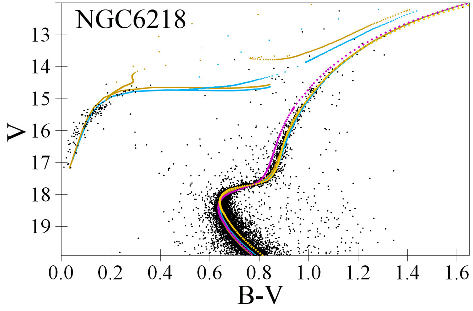}
\caption{The same as Fig.~\ref{cmd01} but for $B-V$ versus $V$ CMD of NGC\,6218 based on the data from \citetalias{stetson2019}. 
The isochrones are calculated with the best-fitting parameters from Table~\ref{ngc6218results}. 
}
\label{cmd07}
\end{figure}

\begin{figure}
\includegraphics{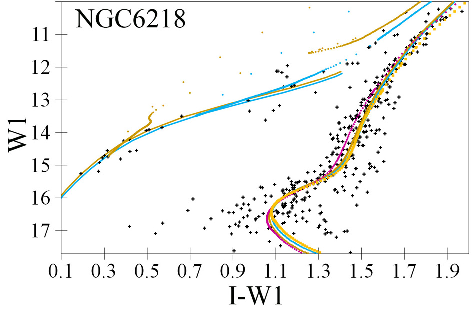}
\caption{The same as Fig.~\ref{cmd01} but for $I-W1$ versus $W1$ CMD of NGC\,6218 based on the data from \citetalias{stetson2019} and unWISE. 
The isochrones are calculated with the best-fitting parameters from Table~\ref{ngc6218results}. 
}
\label{cmd08}
\end{figure}

\section{Results}
\label{results}

Using this wealth of photometric data allows us to fit isochrones to dozens of CMDs with different colours.
As in \citetalias{ngc5904} and \citetalias{ngc6205}, our results for different CMDs are consistent within their precision.
Therefore, we show only some examples of the CMDs with isochrone fits in Figs~\ref{cmd01}--\ref{cmd08}
\footnote{In all the CMDs, the colour is the abscissa and the magnitude in the redder filter is the ordinate.}, 
while the derived ages, distances, and reddenings for NGC\,288, NGC\,362, and NGC\,6218 are presented in Tables~\ref{ngc288results}, 
\ref{ngc362results}, and \ref{ngc6218results}, respectively.
Figures for all the CMDs can be provided on request.

The isochrone fitting fails for some UV CMDs in this study, similar to NGC\,6205 in \citetalias{ngc6205}.
Hence, such UV CMDs are not presented in Tables~\ref{ngc288results}--\ref{ngc6218results} and not used for our final results.

In order to indicate the difference between the primordial and helium-enriched isochrones used, 
both the isochrones are shown in Figs~\ref{cmd01}--\ref{cmd08}. The interpolated isochrone for $Y=0.26$ is between them, as discussed in Sect.~\ref{iso},
and it is not shown for clarity.
Thus, in Figs~\ref{cmd01}--\ref{cmd08} we show four isochrones for the RGB, SGB, TO, and MS and two BaSTI-IAC isochrones for the HB and AGB.
For most CMDs, especially for those in the optical range, the isochrone-to-fiducial fitting is so precise that the best-fitting isochrones of 
different models almost coincide with each other and with our fiducial sequence on the scales of our CMD figures for the RGB, SGB, and TO.
Therefore, we do not show our fiducial sequences in Figs~\ref{cmd01}--\ref{cmd08}.

The following are notes on individual CMDs in Figs~\ref{cmd01}--\ref{cmd08}.

Fig.~\ref{cmd01} presents the {\it Gaia} EDR3 dataset. It is about a magnitude deeper than that of {\it Gaia} DR2 and contains much more stars.
We compare the magnitudes and colours of the common stars of the EDR3 and DR2 datasets and find no significant systematic difference. 
Therefore, we show both the BaSTI-IAC and
DSEP isochrones in Fig.~\ref{cmd01}, although the former fits the {\it Gaia} EDR3 dataset, while the latter -- the {\it Gaia} DR2 dataset.
An agreement of the BaSTI-IAC and DSEP isochrones is evident from Fig.~\ref{cmd01}.

Figs~\ref{cmd01}, \ref{cmd02}, and some other CMDs show that:
\begin{enumerate}
\item The tip of the RGB is mainly populated by stars with the extreme helium enrichment (see \citealt{savino2018}).
\item There is a clear segregation of the two populations at the AGB for the old GCs (NGC\,288 and NGC\,6218).
\item The BaSTI-IAC MS is too blue \citep{pietrinferni2021}. This leads to a conflict between the two proxies of age, the SGB length and 
the HB--SGB magnitude difference: usually the former is too long w.r.t. the latter.
\item There is a colour offset of the isochrones from the bluest HB stars of the old GCs. 
This offset cannot be eliminated by any reasonable variation of distance, age, or reddening.
This offset can be explained by a special evolution of the bluest HB stars and/or by their unusual helium abundance \citep{heber2016}.
\item The length of the HB suggests the masses for the majority of the HB stars within $0.54-0.62$, $0.63-0.80$, and $0.54-0.62$ $M_{\sun}$ for 
NGC\,288, NGC\,362, and NGC\,6218, respectively. However, both the {\it Gaia} EDR3 photometry in all three filters and SMSS photometry 
in the $g_\mathrm{SMSS}$, $r_\mathrm{SMSS}$, and $i_\mathrm{SMSS}$ filters show few interesting marginals at the ZAHB:
the reddest HB stars of NGC\,288 with probable exceptionally high mass of 0.78 $M_{\sun}$ - 
{\it Gaia} EDR3 2342901537529620352 and {\it Gaia} EDR3 2342907992863854208,
the bluest HB stars of NGC\,362 with probable exceptionally low mass of 0.6 $M_{\sun}$ - 
{\it Gaia} EDR3 4690839418131927680 and {\it Gaia} EDR3 4690886353541485824,
and the reddest HB stars of NGC\,6218 with probable exceptionally high mass of 0.72 $M_{\sun}$ - 
{\it Gaia} EDR3 4379077294127099264 and {\it Gaia} EDR3 4379075984156919168.
For the first time, very accurate {\it Gaia} EDR3 PMs ensure a very high probability for these stars to be cluster members.
Properties of these marginals should be investigated in further studies.
\end{enumerate}

Our cleaning of the datasets may differ from that fulfilled by their authors. Hence, our fiducial sequences may differ from those derived by the authors.
For example, Fig.~\ref{cmd04} shows that the fiducial sequence by \citet{bernard2014} is a good representation of the MS and SGB of our sample, 
but is about 0.02 mag bluer than its RGB and HB. The reason for this discrepancy is a deeper cleaning of the Pan-STARRS dataset in our study.
However, usually our fiducial sequence agrees with that derived by the dataset authors.
An example is presented in Fig.~\ref{cmd05}:
the fiducial sequence by \citetalias{bellazzini2001} coincides with our fiducial sequence, which is not shown because it almost coincides with 
the best-fitting isochrones.

Fig.~\ref{cmd06} shows a rare CMD with a significant colour difference [$\Delta(F336W-F438W)\approx0.07$ mag] between the populations at the RGB.
This is a CMD for the UV filters. It is seen that two DSEP isochrones successfully fit the two observed RGBs.
DSEP suggests that the primordial RGB population is redder than the helium-enriched one.
Unlike DSEP, the BaSTI-IAC isochrones for the different populations almost coincide at the RGB. In such a case, both best-fitting isochrones should be
exactly between the observed RGBs, if one takes into account the RGBs only. However, the best-fitting of the HB, AGB, SGB, and MS forces the best-fitting
RGB isochrones to move blueward.

Fig.~\ref{cmd08} shows a typical optical--IR CMD. Usually such a CMD contains less stars, more contaminants, and less pronounced HB and MS than 
an optical CMD. However, such a CMD still provides reliable distance, age and reddening estimates.

The predicted uncertainties of the derived distance, age, and reddening are very similar to those described in appendix A of \citetalias{ngc6205}.
For a typical CMD these uncertainties are about 5 per cent, 7 per cent and 0.03 mag for the derived distance, age, and reddening, respectively.

For each combination of a CMD/fiducial sequence and its best-fitting model/isochrone we find the maximal offset of this isochrone w.r.t. 
this fiducial sequence along the reddening vector (i.e. nearly along the colour). 
Such an offset is presented in Tables~\ref{ngc288results}--\ref{ngc6218results} after each value of reddening as its empirical uncertainty.
We consider cases with an isochrone-to-fiducial offset greater than $0.15$ mag as a fitting failure for the corresponding CMD.

\begin{figure*}
\includegraphics{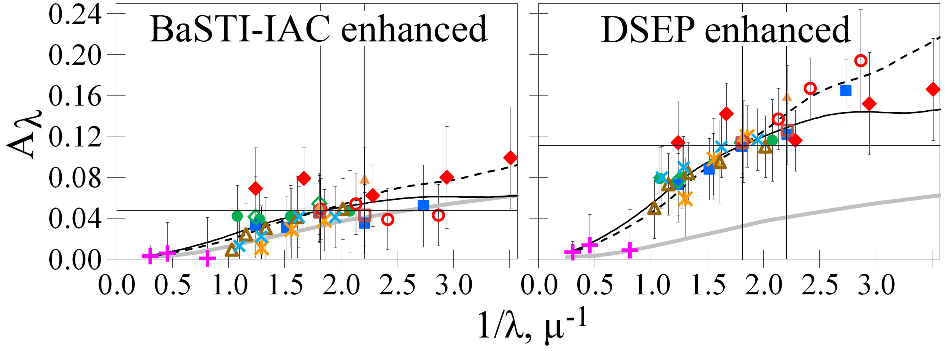}
\caption{The empirical extinction laws for NGC\,288 from the isochrone fitting by the different models.
The datasets are:
{\it HST} ACS and WFC3 by \citet{nardiello2018} -- red diamonds;
{\it Gaia} -- yellow snowflakes;
\citetalias{stetson2019} -- blue squares;
\citetalias{bellazzini2001} -- open green diamonds;
\citet{bolte1992} -- yellow triangles;
\citetalias{sollima2016} -- open brown squares;
\citetalias{grundahl1999} -- open red circles;
Pan-STARRS -- open brown triangles;
DES -- green circles;
SMSS -- blue inclined crosses;
IR datasets by \citet{davidge1997} and unWISE -- purple upright crosses.
For the $B$ and $V$ filters, which are denoted by the vertical lines, the symbols are almost overlapping.
The thick grey curve shows the extinction law of \citet{schlafly2016} with $A_\mathrm{V}=0.033$ mag.
The black dotted and solid curves show the extinction law of \citetalias{ccm89} with $R_\mathrm{V}=3.1$ and $5.7$, respectively,
with the derived $A_\mathrm{V}$, which is shown by the horizontal line.
}
\label{ngc288law}
\end{figure*}

\begin{figure*}
\includegraphics{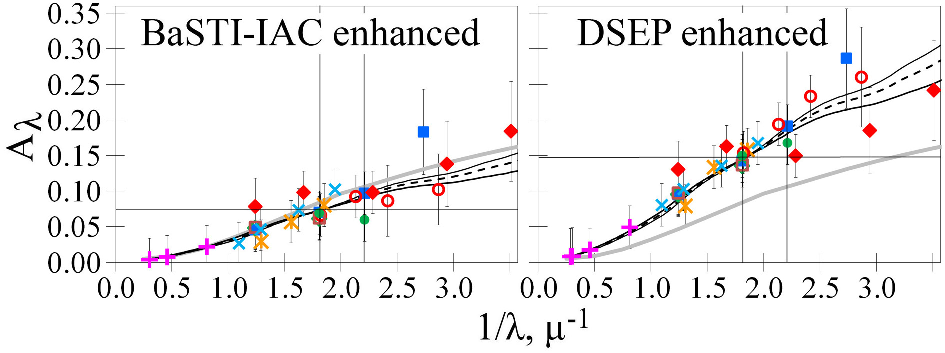}
\caption{The same as Fig.~\ref{ngc288law} but for NGC\,362.
The datasets are:
{\it HST} ACS and WFC3 -- red diamonds;
{\it HST} WFPC2 by \citet{piotto2002} -- green circles;
{\it Gaia} -- yellow snowflakes;
MCPS -- blue squares;
\citetalias{bellazzini2001} -- open green diamonds;
\citetalias{rosenberg2000} -- open brown squares;
\citetalias{grundahl1999} -- open red circles;
SMSS -- blue inclined crosses;
IR datasets by \citet{cohen2015}, unWISE, and SAGE -- purple upright crosses.
For the $B$ and $V$ filters, which are denoted by the vertical lines, the symbols are almost overlapping.
The thick grey curve shows the extinction law of \citet{schlafly2016} with $A_\mathrm{V}=0.086$ mag.
The black thin, dotted and thick curves show the extinction law of \citetalias{ccm89} with $R_\mathrm{V}=2.8$, $3.1$ and $3.6$, respectively, 
with the derived $A_\mathrm{V}$, which is shown by the horizontal line.
}
\label{ngc362law}
\end{figure*}

\begin{figure*}
\includegraphics{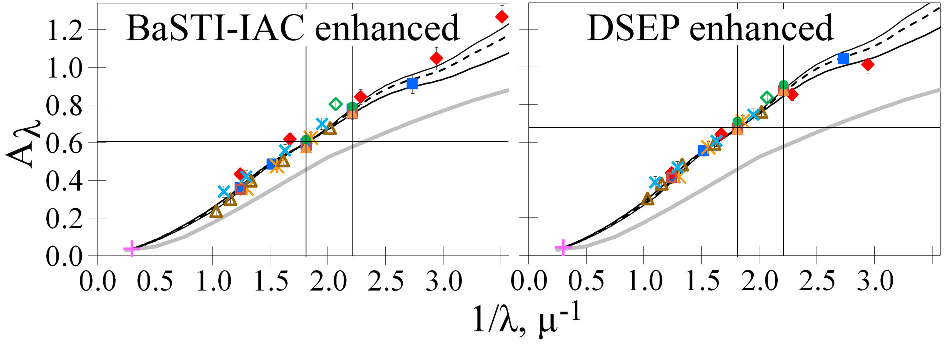}
\caption{The same as Fig.~\ref{ngc288law} but for NGC\,6218.
The datasets are:
{\it HST} ACS and WFC3 by \citet{nardiello2018} -- red diamonds;
{\it HST} ACS by \citet{simioni2018} -- open green diamonds;
{\it HST} WFPC2 by \citet{piotto2002} -- green circles;
{\it Gaia} -- yellow snowflakes;
\citetalias{stetson2019} -- blue squares;
\citetalias{zloczewski2012} -- yellow triangles;
\citetalias{hargis2004} -- open brown squares;
Pan-STARRS -- open brown triangles;
SMSS -- blue inclined crosses;
unWISE -- purple upright crosses.
For the $B$ and $V$ filters, which are denoted by the vertical lines, the symbols are almost overlapping.
The thick grey curve shows the extinction law of \citet{schlafly2016} with $A_\mathrm{V}=0.467$ mag.
The black thin, dotted and thick curves show the extinction law of \citetalias{ccm89} with $R_\mathrm{V}=2.9$, $3.1$ and $3.5$, respectively, with 
the derived $A_\mathrm{V}$, which is shown by the horizontal line.
The error bars are typically shorter than the size of the symbols used.
}
\label{ngc6218law}
\end{figure*}

For most CMDs, the predicted and empirical uncertainties are comparable. The largest value in such a pair of the uncertainties is shown by 
an error bar in Figs~\ref{ngc288law}--\ref{ngc6218law}, which demonstrate the resulting extinction laws.

\subsection{Reddening and extinction}
\label{redext}

Each CMD provides us with independent estimates of age, distance, and reddening.
Each dataset in several filters, providing several CMDs, gives us an independent set of the derived reddenings.
With such a set of reddenings, we can calculate the extinction for each filter of the dataset, if we adopt an extinction at the dataset filter 
with the longest wavelength, i.e. extinction zero-point.

Extinction and its uncertainty are minimal for IR filters. Therefore, we cross-identify each optical dataset with an IR dataset and adopt the 
extinction in an IR filter as the extinction zero-point for the optical dataset.
The exceptions are made for three datasets, which cannot be cross-identified with an IR dataset and, hence, they have no IR extinction zero-point:
the photometry from \citet{piotto2002} and both the datasets represented by fiducial sequences only, without data for individual stars 
(\citealt{bolte1992} and \citealt{davidge1997}).

Details on calculating extinctions and empirical extinction law are provided in section 5.1 of \citetalias{ngc6205}.
Here we briefly summarise some important points: (i) we initially adopt any reasonable extinction in an IR filter used, 
(ii) we calculate extinctions in all the filters of the dataset by use of the corresponding reddenings and the IR extinction, 
(iii) the derived extinctions draw up an empirical extinction law for this specific dataset,
(iv) we recalculate the IR extinction with the derived extinction law,
and then repeat the steps (ii)--(iv) iteratively.
We need only a few iterations to converge, since any reasonable variation of the extinction law results in a little variation of the IR 
extinction. For the clusters under consideration this variation is $<0.01$ mag.

We use the unWISE IR photometry for all the GCs, while for NGC\,362 we also use the IR photometry from \citet{cohen2015} and SAGE (except for 
BaSTI-IAC giving no SAGE isochrone).
Both the $J$ and $K_s$ data from \citet{cohen2015} are used in order to suppress possible systematics due to possible poor determination of the IR 
filter profiles.

Owing to the IR extinction zero-points, all the derived extinctions are rather precise being determined on a long UV--IR or optical--IR 
wavelength baseline. For example, we calculate
\begin{equation}
\label{avaw1}
A_\mathrm{V}=(A_\mathrm{V}-A_\mathrm{W1})+A_\mathrm{W1}=E(V-W1)+A_\mathrm{W1}
\end{equation}
from the reddening $E(V-W1)$ and the IR extinction $A_\mathrm{W1}$.

We adjust the datasets based on some CMDs obtained with pairs of filters from different datasets but of similar effective wavelengths.
This adjustment follows its detailed description in section 6 of \citetalias{ngc6205}.
We adjust magnitudes and colours from different datasets in order to minimise the scatter of extinctions from the datasets around an 
average extinction for each model.
This adjustment does not change the average extinctions, distances, and ages derived for each model.
All the adjustment corrections are within $\pm0.035$ mag\footnote{The unWISE--SAGE $W1-3.6~\mu m$ colour differences for the common NGC\,362 
stars show that the DSEP predictions for SAGE should be corrected by $+0.12$ mag.}.

The adjustment corrections to the same model--dataset pair (such as BaSTI-IAC--{\it Gaia} EDR3) appear slightly different for different GCs.
Hence, they may be due to some systematic errors of the datasets, which vary from one GC to another with observational conditions, 
celestial position, or time (see a detailed analysis by \citealt{anderson2008}).

Some datasets cannot be adjusted since they have no appropriate filters, or they have little, if any, common stars with the other datasets.
An example is the {\it HST} ACS/WFC3 datasets of \citet{nardiello2018}.
They provide a systematically higher extinction (red diamonds in Figs~\ref{ngc288law}--\ref{ngc6218law}) w.r.t. the other datasets.
However, the {\it HST} ACS/WFC3 datasets are not adjusted w.r.t. the other datasets, since the former are obtained within approximately 2 
arcmin from the cluster centres, where the other datasets have little stars.
Therefore, we cannot make a final conclusion about the reason for this deviation of the {\it HST} ACS/WFC3 extinction estimates.
This may be due to some systematic errors of these datasets or due to a population variation in the cluster centres.
This deviation of the {\it HST} ACS/WFC3 extinction estimates needs further study.

\begin{table*}
\def\baselinestretch{1}\normalsize\normalsize
\caption[]{The estimates of $A_\mathrm{V}$ for NGC\,288 from the various datasets and models.
}
\label{ngc288av}
\[
\begin{tabular}{lcc}
\hline
\noalign{\smallskip}
 & \multicolumn{2}{c}{Dataset} \\
\noalign{\smallskip}
Model & \citetalias{stetson2019}, \citetalias{grundahl1999}, \citetalias{bellazzini2001}, \citetalias{sollima2016} & The same plus {\it HST}, PS1, DES, {\it Gaia}, and SMSS \\
\hline
\noalign{\smallskip}
BaSTI-IAC enhanced     & $0.049\pm0.004$     & $0.047\pm0.009$  \\    
DSEP enhanced          & $0.112\pm0.003$     & $0.111\pm0.006$  \\    
\noalign{\smallskip}
Mid-range value        & $0.080\pm0.032$     & $0.079\pm0.032$ \\   
\hline
\end{tabular}
\]
\end{table*}

\begin{table*}
\def\baselinestretch{1}\normalsize\normalsize
\caption[]{The estimates of $A_\mathrm{V}$ for NGC\,362 from the various datasets and models.
}
\label{ngc362av}
\[
\begin{tabular}{lcc}
\hline
\noalign{\smallskip}
 & \multicolumn{2}{c}{Dataset} \\
\noalign{\smallskip}
Model & MCPS, \citetalias{grundahl1999}, \citetalias{rosenberg2000}, \citetalias{bellazzini2001} & The same plus {\it HST} and {\it Gaia} \\
\hline
\noalign{\smallskip}
BaSTI-IAC enhanced     & $0.065\pm0.005$    & $0.073\pm0.011$   \\
DSEP enhanced          & $0.142\pm0.009$    & $0.146\pm0.010$   \\
\noalign{\smallskip}
Mid-range value        & $0.104\pm0.039$     & $0.109\pm0.037$ \\
\hline
\end{tabular}
\]
\end{table*}

\begin{table*}
\def\baselinestretch{1}\normalsize\normalsize
\caption[]{The estimates of $A_\mathrm{V}$ for NGC\,6218 from the various datasets and models.
}
\label{ngc6218av}
\[
\begin{tabular}{lcc}
\hline
\noalign{\smallskip}
 & \multicolumn{2}{c}{Dataset} \\
\noalign{\smallskip}
Model & \citetalias{stetson2019}, \citetalias{zloczewski2012}, \citetalias{hargis2004} & The same plus {\it HST}, Pan-STARRS, and {\it Gaia} \\
\hline
\noalign{\smallskip}
BaSTI-IAC enhanced     & $0.583\pm0.009$    & $0.600\pm0.029$ \\
DSEP enhanced          & $0.659\pm0.019$    & $0.668\pm0.017$  \\
\noalign{\smallskip}
Mid-range value        & $0.621\pm0.038$    & $0.634\pm0.034$ \\
\hline
\end{tabular}
\]
\end{table*}

The extinction $A_\mathrm{V}$ requires special attention, since we have a wealth of data to determine it.
Moreover, we can use some datasets without the $V$ filter to validate the average system obtained for the datasets with the $BVI$ filters.
The derived extinctions $A_\mathrm{V}$ are presented in Tables~\ref{ngc288av}, \ref{ngc362av}, and \ref{ngc6218av} for 
NGC\,288, NGC\,362, and NGC\,6218, respectively.
In these tables, the datasets, listed in the central column, allow a direct determination of $A_\mathrm{V}$ by use of eq.~(\ref{avaw1}) or a similar 
equation for another IR filter, without knowledge of the extinction law.
In the rightmost column we present the results obtained with these and some additional datasets.
Each additional dataset includes several filters and has an IR extinction zero-point, but it provides no direct determination of $A_\mathrm{V}$.
In this case, we calculate $A_\mathrm{V}$ from the extinction in adjacent filters by use of the following extinction law relations:
\begin{enumerate}
\item $A_\mathrm{V}=0.53(A_\mathrm{G_\mathrm{BP}}+A_\mathrm{G})$ for {\it Gaia},
\item $A_\mathrm{V}=0.49(A_\mathrm{g_\mathrm{DECam}}+A_\mathrm{r_\mathrm{DECam}})$ for DES,
\item $A_\mathrm{V}=0.50(A_\mathrm{g_\mathrm{PS1}}+A_\mathrm{r_\mathrm{PS1}})$ for Pan-STARRS,
\item $A_\mathrm{V}=0.44(A_\mathrm{F438W}+A_\mathrm{F606W})$ for {\it HST}, and
\item $A_\mathrm{V}=0.51(A_\mathrm{g_\mathrm{SMSS}}+A_\mathrm{r_\mathrm{SMSS}})$ for SMSS.
\end{enumerate}
The coefficients in these relations are rather close to 0.5, since we use the pairs of filters, which are rather close and nearly symmetric in 
wavelength w.r.t. the $V$ filter. These coefficients vary only within $\pm0.015$ for the extinction laws of \citet{fitzpatrick1999}, 
\citet{schlafly2016}, \citet{wang2019}, and \citetalias{ccm89} with $2.4<R_\mathrm{V}<4.1$. 
Therefore, the use of these coefficients without knowing the proper extinction law provides a relative uncertainty of the obtained $A_\mathrm{V}$ of 
less than $0.015/0.44=0.034$ and, hence, a negligible contribution of $\pm0.004$, $\pm0.004$, and $\pm0.02$ mag to the final $A_\mathrm{V}$ 
uncertainty in Tables~\ref{ngc288av}--\ref{ngc6218av}.

The $A_\mathrm{V}$ estimates in Tables~\ref{ngc288av}--\ref{ngc6218av} agree for the different datasets (i.e. those in the central and rightmost columns). 
Hereafter, we prefer the values obtained for the larger lists of the datasets.

However, the $A_\mathrm{V}$ estimates in Tables~\ref{ngc288av}--\ref{ngc6218av} disagree for the different models (i.e. those in different rows).
This disagreement seems to be due to a systematic difference between the models.
We decide to calculate our final estimates as the mid-range (a middle value between the model estimates) values\footnote{Here the mid-range values 
are equal to the mean values. However, we entitle designate the bottom rows of Tables~\ref{ngc288av}--\ref{ngc6218av} as mid-range values in order to 
emphasise that their uncertainties are not standard deviations but half the differences between the DSEP and BaSTI-IAC estimates.} 
with their uncertainties as half the differences between the DSEP and BaSTI-IAC estimates:
$A_\mathrm{V}=0.08\pm0.03$, $0.11\pm0.04$, and $0.63\pm0.03$ mag for NGC\,288, NGC\,362, and NGC\,6218, respectively.

Following Tables~\ref{ngc288av}--\ref{ngc6218av}, Figs~\ref{ngc288law}--\ref{ngc6218law} show a low scatter for the datasets, while a large 
difference between the models.

Our $A_\mathrm{V}$ estimates for NGC\,288 and NGC\,6218 are considerably higher than those in Table~\ref{properties} from the 2D dust emission 
maps of \citet{sfd98}, \citet{schlaflyfinkbeiner2011}, and \citet{2015ApJ...798...88M}.
This is also seen in Figs~\ref{ngc288law}--\ref{ngc6218law} where the thick grey curves show the extinction law 
of \citet{schlafly2016} with the extinction estimate from \citet{schlaflyfinkbeiner2011}:
DSEP for all the GCs and both the models for NGC\,6218 provide the empirical extinction laws, which cannot be reconciled with the prediction 
of \citet{schlaflyfinkbeiner2011}.

However, our $A_\mathrm{V}$ estimates are lower than those from \citet{wagner2016,wagner2017}, who used a Bayesian single- and two-population analysis 
for the {\it HST} ACS data and obtained
$0.084<A_\mathrm{V}<0.140$, $0.117<A_\mathrm{V}<0.128$, and $0.682<A_\mathrm{V}<0.695$ for NGC\,288, NGC\,362, and NGC\,6218, respectively.

This wide spread of the $A_\mathrm{V}$ estimates from the literature may be due to the incorrect extinction-to-reddening coefficients used. 
Usually, an extinction estimate is extrapolated from a reddening, measured in a narrow wavelength range, 
by use of an adopted extinction law. If such a law differs from the proper law, this introduces an additional extrapolation error. 
The most common way (giving all the estimates of the 2D maps in Table~\ref{properties}) has been the extrapolation of the observed reddening 
$E(B-V)$ to $A_\mathrm{V}$. Therefore, it is interesting to calculate and compare $E(B-V)$ estimates for these GCs.

To estimate $E(B-V)$ from Tables~\ref{ngc288results}--\ref{ngc6218results}, we use both the $E(B-V)$ estimates themselves and
the $E(b-y)$ reddenings in the adjacent Str\"omgren $b$ and $y$ filters by use of the extinction law relation 
$E(B-V)=1.29E(b-y)$\footnote{{\it HST} WFC3 $F438W$ and ACS $F606W$ filter pair also may be a good proxy of the $B$ and $V$ filters. 
However, belonging to the different detectors (WFC3 and ACS), these filters may allow a small instrumental systematic error in the $F438W-F606W$ colour.
Even being 0.01 mag, such an error significantly affects very small reddening $E(F438W-F606W)$ of NGC\,288 and NGC\,362. Therefore, we decide not to use 
$E(F438W-F606W)$ in our $E(B-V)$ calculation.}.
Since this pair of filters is rather close and nearly symmetric in wavelength w.r.t. the $BV$ pair, the coefficient in this relation varies 
negligibly (only within $\pm0.01$) for the extinction laws of \citet{fitzpatrick1999}, \citet{schlafly2016}, \citet{wang2019}, and \citetalias{ccm89} 
with $2.3<R_\mathrm{V}<4.0$.

We obtain $E(B-V)=0.014\pm0.010$, $0.028\pm0.011$, and $0.189\pm0.010$ mag for NGC\,288, NGC\,362, and NGC\,6218, respectively.
Similar to the $A_\mathrm{V}$ estimates, the DSEP and BaSTI-IAC $E(B-V)$ estimates differ significantly. 
Hence, these uncertainties are half the differences between the model estimates.
These $E(B-V)$ estimates for NGC\,288 and NGC\,362 are about their lowest reddening estimates in Table~\ref{properties}. 
In combination with rather high $A_\mathrm{V}$ estimates, this suggests that the observed extinction law may deviate from the laws adopted for the
estimates in Table~\ref{properties}.
This is seen in Figs~\ref{ngc288law} and \ref{ngc362law} with the empirical extinction laws for NGC\,288 and NGC\,362, respectively:
all the data in the $BV$ range show little increase of extinction with the wavelength.

Given our estimates of $A_\mathrm{V}$ and $E(B-V)$ with their uncertainties, our estimates of $R_\mathrm{V}$ from the different models are consistent, 
although they are poorly constrained for NGC\,288 and NGC\,362:
$R_\mathrm{V}>3.1$ for NGC\,288, $R_\mathrm{V}>2.1$ for NGC\,362, and $R_\mathrm{V}=3.35^{+0.25}_{-0.23}$ for NGC\,6218.
We emphasise that these $R_\mathrm{V}$ estimates characterise the empirical extinction law only between the $B$ and $V$ filters.
Alternatively, $R_\mathrm{V}$ for the whole wavelength range under consideration can be estimated from the derived extinctions 
in all the filters of all the datasets, separately for each model. This is the $R_\mathrm{V}$ estimate, which we use to calculate the IR extinction 
zero-points mentioned above.
In Figs~\ref{ngc288law}--\ref{ngc6218law} we depict the \citetalias{ccm89} extinction law with the derived $A_\mathrm{V}$ and 
various $R_\mathrm{V}$ shown by the black curves.
Taking the error bars into account, all our results are in acceptable agreement with the common \citetalias{ccm89} extinction law with 
$R_\mathrm{V}=3.1$, which is shown by the dotted curve.
Hence, we adopt this law to calculate the IR extinction zero-points.
However, Figs~\ref{ngc288law}--\ref{ngc6218law} show that the empirical extinction law can be better presented by the 
\citetalias{ccm89} law with $3.1<R_\mathrm{V}<5.7$, $2.8<R_\mathrm{V}<3.6$, and $2.9<R_\mathrm{V}<3.5$ for NGC\,288, NGC\,362, and NGC\,6218, respectively.

\citetalias{ngc6205} has shown that all most reliable extinction laws, such as those of \citet{fitzpatrick1999}, \citet{schlafly2016}, 
and \citet{wang2019}, with the same $A_\mathrm{V}$ are very close to the \citetalias{ccm89} extinction law and to each other.
Therefore, the choice of law or $R_\mathrm{V}$ can barely change the agreement of a law with our results in Figs~\ref{ngc288law}--\ref{ngc6218law}, 
but the choice of $A_\mathrm{V}$ can.
Thus, we emphasise that Tables~\ref{ngc288av}--\ref{ngc6218av} contain our most important results about extinction for these GCs.

The reddening or extinction estimates from \citet{sfd98}, \citet{schlaflyfinkbeiner2011}, and \citet{2015ApJ...798...88M} in Table~\ref{properties}
are most frequently used in studies of extragalactic objects at middle and high Galactic latitudes.
A possible underestimation of low extinctions by these studies has been found by various methods \citep{gm2017,gm2017big,gm2018,gm2021}, 
in particular, in \citetalias{ngc5904} and \citetalias{ngc6205} for both NGC\,5904 and NGC\,6205. 
However, this study shows that this issue is even more complex, when extinction is underestimated while reddening is overestimated by these maps 
for some GCs. 
This suggests that a proper extinction law differs from an adopted law.
In the case of an uncertain proper extinction law, we recommend to use extinction instead of reddening, since extinction is valid on a 
longer wavelength baseline.

\begin{table*}
\def\baselinestretch{1}\normalsize\normalsize
\caption[]{Our age (Gyr) and distance (kpc) estimates for NGC\,288.
}
\label{ngc288agedist}
\[
\begin{tabular}{lccc}
\hline
\noalign{\smallskip}
                                          & DSEP            & BaSTI-IAC        &  Mean value \\
\hline
\noalign{\smallskip}
Mean distance for $0.42<\lambda<1.0$ nm   & $9.04\pm0.23$   &   $8.89\pm0.16$  &  $8.96\pm0.21$ \\
Mean distance for the UV and IR range     & $8.78\pm0.41$   &   $8.79\pm0.32$  &  $8.79\pm0.36$ \\
\noalign{\smallskip}
Mean age for $0.42<\lambda<1.0$ nm        & $12.45\pm0.60$  &   $14.63\pm0.92$ &  $13.54\pm1.34$ \\
Mean age for the UV and IR range          & $12.53\pm0.78$  &   $13.81\pm0.85$ &  $13.17\pm1.04$ \\
\hline
\end{tabular}
\]
\end{table*}


\begin{table*}
\def\baselinestretch{1}\normalsize\normalsize
\caption[]{Our age (Gyr) and distance (kpc) estimates for NGC\,362.
}
\label{ngc362agedist}
\[
\begin{tabular}{lccc}
\hline
\noalign{\smallskip}
                                            & DSEP            & BaSTI-IAC         & Mean value \\
\hline
\noalign{\smallskip}
Mean distance for $0.42<\lambda<1.0$ nm    & $8.95\pm0.27$   &   $9.02\pm0.13$   &  $8.98\pm0.21$  \\
Mean distance for the UV and IR range      & $8.86\pm0.34$   &   $8.99\pm0.21$   &  $8.92\pm0.29$ \\
\noalign{\smallskip}
Mean age for $0.42<\lambda<1.0$ nm         & $10.42\pm0.47$  &   $11.54\pm0.54$    & $10.98\pm0.76$ \\
Mean age for the UV and IR range           & $10.63\pm1.16$  &   $11.22\pm0.93$    & $10.89\pm1.10$ \\
\hline
\end{tabular}
\]
\end{table*}


\begin{table*}
\def\baselinestretch{1}\normalsize\normalsize
\caption[]{Our age (Gyr) and distance (kpc) estimates for NGC\,6218.
}
\label{ngc6218agedist}
\[
\begin{tabular}{lccc}
\hline
\noalign{\smallskip}
                                          &  DSEP            &   BaSTI-IAC      &  Mean value \\
\hline
\noalign{\smallskip}
Mean distance for $0.42<\lambda<1.0$ nm   & $5.09\pm0.24$    &   $5.00\pm0.17$  &  $5.04\pm0.21$   \\
Mean distance for the UV and IR range     & $5.11\pm0.19$    &   $5.12\pm0.19$  &  $5.11\pm0.18$   \\
\noalign{\smallskip}
Mean age for $0.42<\lambda<1.0$ nm        & $12.75\pm0.73$   &   $14.89\pm0.99$ &  $13.82\pm1.38$ \\
Mean age for the UV and IR range          & $13.25\pm0.72$   &   $14.14\pm0.78$ &  $13.71\pm0.86$ \\
\hline
\end{tabular}
\]
\end{table*}


\subsection{Age and distance}
\label{agedist}

Our age and distance estimates for NGC\,288, NGC\,362, and NGC\,6218 are presented in Tables~\ref{ngc288agedist}, \ref{ngc362agedist}, and 
\ref{ngc6218agedist}, respectively.
Their uncertainties are standard deviations calculated from all reliable independent CMDs with the given model applied.

We separate the results for the optical range ($0.42<\lambda<1.0$ nm) and the UV and IR ranges.
Similar to NGC\,6205 in \citetalias{ngc6205}, the results for the ranges differ, both systematically and by their standard deviations.
Assuming that the models are more accurate in the optical range, we decide to use the estimates in the optical range as the final, most probable estimates.

The right column presents the mean value and standard deviation of the combuned results for the two models.
It is seen that the models provide consistent estimates of distance, but inconsistent estimates of age.
Therefore, we accept the final uncertainty of our distance estimates as the standard deviation of the combined sample divided by the square root of 
the number of the CMDs used.
In contrast, we accept the final uncertainty of our age estimates as half the difference between the model estimates.

Our final estimates for NGC\,288, NGC\,362, and NGC\,6218, respectively, are:
\begin{enumerate}
\item age is $13.5\pm1.1$, $11.0\pm0.6$, and $13.8\pm1.1$ Gyr, 
\item distance is $8.96\pm0.05$, $8.98\pm0.06$, and $5.04\pm0.05$ kpc,
\item distance modulus $(m-M)_0$ is $14.76\pm0.01$, $14.77\pm0.01$, and $13.51\pm0.02$ mag,
\item apparent $V$-band distance modulus $(m-M)_\mathrm{V}$ is $14.84\pm0.03$, $14.88\pm0.04$, and $14.14\pm0.04$ mag.
\end{enumerate}

Our distance estimates agree with those from the recent compilation of all distance determinations by \citet{baumgardt2021} 
within $0.2\sigma$, $0.9\sigma$, and $0.7\sigma$ of the stated uncertainties for NGC\,288, NGC\,362, and NGC\,6218, respectively. 
It is worth noting that other studies in the compilation of \citet{baumgardt2021} have considered only one or a few datasets, while in our paper 
a dozen of datasets for each cluster is analysed in detail. Therefore, our distance estimates for these clusters should be among most accurate.

Our distances correspond to the parallaxes $0.1116\pm0.0006$, $0.1114\pm0.0008$, and $0.1984\pm0.0020$ mas
for NGC\,288, NGC\,362, and NGC\,6218, respectively.
They agree within their uncertainties with the median corrected {\it Gaia} EDR3 parallaxes, which we obtain in Sect.~\ref{edr3} and present
in Table~\ref{properties}:
$0.114\pm0.011$, $0.119\pm0.011$, and $0.210\pm0.011$ mas for NGC\,288, NGC\,362, and NGC\,6218, respectively.
However, all the {\it Gaia} EDR3 parallaxes are higher than ours.
This comparison confirms the conclusion of \citetalias{vasiliev2021} that the corrected {\it Gaia} EDR3 parallaxes are overestimated by 
$\sim0.006-0.009$ mas.
Anyway, it is evident that the GC distances from the {\it Gaia} EDR3 parallaxes are still less accurate than those obtained from such an isochrone fitting.

\begin{figure}
\includegraphics{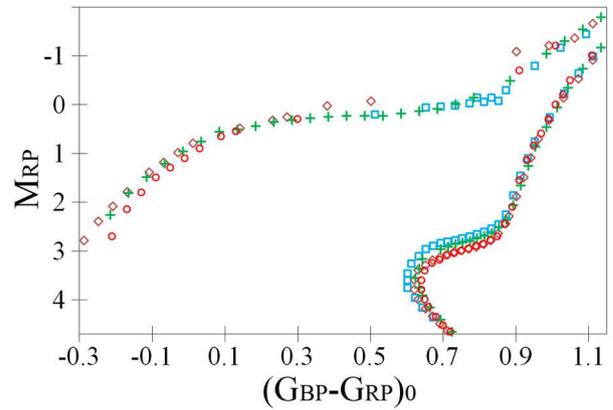}
\caption{The CMD with the {\it Gaia} EDR3 fiducial sequences for NGC\,288 (brown diamonds), NGC\,362 (blue squares), NGC\,5904 (green crosses),
and NGC\,6218 (red circles).
}
\label{hb}
\end{figure}

Our age estimates are within a wide variety of age estimates for these GCs from the literature\footnote{This variety is a reason why there is
no consensus about age as the second parameter.}.
For example, our age estimates are in acceptable agreement with that of $11.5\pm2.0$ Gyr for NGC\,288 from \citet{omalley2017}, as well as with the
age estimates from the isochrone fitting of the {\it HST} ACS data\footnote{We use the same photometry in the $F606W$ and $F814W$ filters.}
by \citet{vandenberg2013}: 
$11.50\pm0.38$, $10.75\pm0.25$, and $13.00\pm0.50$ Gyr for NGC\,288, NGC\,362, and NGC\,6218, respectively.
This age estimate for NGC\,288 is much younger than ours.
However, \citet{vandenberg2013} note for NGC\,288 `Both the slope of the SGB and the location of the RGB are less problematic if the higher age is 
assumed.'

Our age estimates agree with those from the Bayesian single-population analysis of the same {\it HST} ACS data by \citet{wagner2017}: 
$13.50\pm0.01$, $11.46\pm0.10$, and $13.50\pm0.01$ Gyr for NGC\,288, NGC\,362, and NGC\,6218, respectively.
However, our estimates are much less consistent with those from the two-population analysis of the same data by \citet{wagner2016}:
$11.03\pm0.04$, $9.99\pm0.02$, and $12.88\pm0.02$ Gyr for NGC\,288, NGC\,362, and NGC\,6218, respectively.
This may be explained by the fact that two populations of these GCs are not segregated in the $F606W-F814W$ CMD \citep{vandenberg2013}. Hence, 
a two-population analysis seems to be less appropriate in this case.

The models are consistent in their relative age estimates: NGC\,362 is $2.6\pm0.5$ Gyr younger than NGC\,288 and $2.8\pm0.5$ Gyr younger than NGC\,6218
\footnote{These estimates agree with some previous ones, e.g. by \citetalias{bellazzini2001}.}.
These estimates are the mean differences between the age estimates in Tables~\ref{ngc288agedist}--\ref{ngc6218agedist}, computed separately for each model. 
The uncertainties are half the differences between the model estimates.
Following the balance of uncertainties in appendix A of \citetalias{ngc6205}, we note that,
in the case of a relative age calculation, the uncertainties of model, colour--$T_\mathrm{eff}$ relation and bolometric correction are not important. 
Hence, photometric errors (decreased by our use of several datasets) and an uncertainty due to multiple populations dominate uncertainty of the final 
relative age. This explains the much lower uncertainties of the relative ages w.r.t. those of the absolute ages.

NGC\,5904 has a similar metallicity.
Our age estimate $12.15\pm1.0$ Gyr for NGC\,5904 from \citetalias{ngc5904} suggests that it is 1.15 Gyr older than NGC\,362.
Thus, one can see a clear correlation of the derived relative ages with the HB type\footnote{The HB type is defined as $(N_B-N_R)/(N_B+N_V+N_R)$, 
where $N_B$, $N_V$, and $N_R$ are the number of stars that lie blueward of the 
instability strip, the number of RR~Lyrae variables, and the number of stars that lie redward of the instability strip, respectively \citep{lee1994}.}
$+0.98$, $-0.87$, $+0.31$, and $+0.98$ taken from \citet{vandenberg2013} for NGC\,288, NGC\,362, NGC\,5904, and NGC\,6218, respectively.
This is an important result. Since we use by far the most expanded data, our result is a robust confirmation of the long-standing assumption that 
age is the second parameter for these GCs.

A direct comparison of the HBs and SGBs of these GCs is useful. Table~\ref{filters} shows that {\it HST} ACS/WFC3, {\it Gaia}, and SMSS are the only 
common datasets for these GCs.
However, the {\it HST} dataset covers only a few central arcminutes of the fields and shows the unexplained extinction deviation, as noted in 
Sect.~\ref{redext}.
{\it Gaia} EDR3 contains more stars and better defined fiducial sequences than SMSS. Therefore, we compare the HBs in the {\it Gaia} EDR3 CMD.
We transfer our {\it Gaia} EDR3 fiducial sequences into the plane of the dereddened colour $(G_\mathrm{BP}-G_\mathrm{RP})_0$ versus the 
absolute magnitude $M_\mathrm{G_\mathrm{RP}}$ by use of the reddenings and distances from Tables~\ref{ngc288results}--\ref{ngc6218results} and the 
\citetalias{ccm89} extinction law. 
Also, we process the {\it Gaia} EDR3 data of NGC\,5904, obtain its fiducial sequence, and transfer it by use of the reddening and distance 
from \citetalias{ngc5904}.
The fiducial sequences of the four GCs are shown in Fig.~\ref{hb} by different open symbols.
The RGB colours of the different fiducial sequences coincide within $\pm0.01$ mag.
Good agreement between their HB magnitudes within $\pm0.05$ mag is evident, if we compare them w.r.t. the HB of NGC\,5904, which spans the whole 
colour range.
This good agreement\footnote{This agreement is sufficient to see the various HB-SGB magnitude differences for these GCs. 
We ensure the final agreement by use of several datasets.}
for the RGBs and HBs of the different fiducial sequences means that we adopt the correct reddenings (important for positioning RGBs) 
and correct distances (important for positioning HBs).

A small colour offset between the bluest HB fiducial sequence points of NGC\,288, NGC\,5904, and NGC\,6218 remains.
This offset cannot be eliminated, when the HBs and RGBs are fixed (see our discussion of Figs~\ref{cmd01} and \ref{cmd02}).

Fig.~\ref{hb} shows that in the sequence NGC\,362, NGC\,5904, NGC\,288, and NGC\,6218, the HB--SGB magnitude difference increases, 
while the SGB length decreases. Thus, we can infer that the age of the clusters in this sequence increases.
On the other hand, the HB morphology also changes along this sequence:
NGC\,362 has only a red HB, the HB of NGC\,5904 spans the whole colour range, while NGC\,288 and NGC\,6218 have only a blue HB.
Therefore, this correlation of the morphology with the HB--SGB magnitude difference and SGB length strongly suggests the morphology variation with age.

\section{Conclusions}
\label{conclusions}

This study generally follows \citetalias{ngc5904} and \citetalias{ngc6205} in their approach to estimate distance, age, and extinction law 
of Galactic GCs by fitting model isochrones to a multiband photometry. We have considered the famous second-parameter pair NGC\,288 and 
NGC\,362 with a very low extinction, as well as the middle-latitude cluster NGC\,6218 (M12) of nearly the same metallicity. 
For all three clusters, we accepted the metallicity [Fe/H]$=-1.3$ based on the spectroscopy taken from the literature.

We used the photometry of the clusters in many filters from the {\it HST}, unWISE, {\it Gaia} DR2 and EDR3, Pan-STARRS, SkyMapper Southern Sky Survey DR3
and other datasets.
These filters cover a wavelength range from about 235 to 4070\,nm, i.e. from the UV to mid-IR.
Some of the photometric datasets were cross-identified with each other.
This allowed us 
(i) to use an IR photometry with a nearly zero extinction for accurate determination of extinction in all other filters and 
(ii) to estimate some systematic differences of the datasets and suppress them to a level of $<0.01$ mag.

To fit the data, we used the DSEP and BaSTI-IAC theoretical models of the stellar evolution for $\alpha$--enhanced populations of the GCs with different 
helium abundance.
An improvement of these models in recent years makes isochrone-to-data fitting in a typical optical CMD fairly precise.
However, the models are still inconsistent in their predictions for age and reddening.

For NGC\,288, NGC\,362, and NGC\,6218, we derived the most probable 
distances $8.96\pm0.05$, $8.98\pm0.06$, and $5.04\pm0.05$ kpc,
distance moduli $14.76\pm0.01$, $14.77\pm0.01$, and $13.51\pm0.02$ mag,
apparent $V$-band distance moduli $14.84\pm0.03$, $14.88\pm0.04$, and $14.14\pm0.04$ mag,
ages $13.5\pm1.1$, $11.0\pm0.6$, and $13.8\pm1.1$ Gyr, 
extinctions $A_\mathrm{V}=0.08\pm0.03$, $0.11\pm0.04$, and $0.63\pm0.03$ mag, and 
reddenings $E(B-V)=0.014\pm0.010$, $0.028\pm0.011$, and $0.189\pm0.010$ mag, respectively.
This leads to $R_\mathrm{V}=3.35^{+0.25}_{-0.23}$ for NGC\,6218.
The extinction laws for NGC\,288 and NGC\,362 are less certain.

All the models are consistent in their relative age estimates: NGC\,362 is $2.6\pm0.5$ Gyr younger than NGC\,288 and $2.8\pm0.5$ Gyr younger than NGC\,6218.
Using the results from \citetalias{ngc5904}, we find that NGC\,362 is 1.15 Gyr younger than NGC\,5904 with a similar metallicity.
Taking into account the different HB morphology of these four GCs, our findings confirm the long-standing assumption that age is their second parameter.

Based on {\it Gaia} EDR3, we provide the lists of reliable members of the clusters and systemic proper motions with their systematic 
uncertainties in mas\,yr$^{-1}$:
$$\mu_{\alpha}\cos(\delta)=4.147\pm0.024,\; \mu_{\delta}=-5.704\pm0.025$$
$$\mu_{\alpha}\cos(\delta)=6.696\pm0.024,\; \mu_{\delta}=-2.543\pm0.024$$
$$\mu_{\alpha}\cos(\delta)=-0.204\pm0.024,\; \mu_{\delta}=-6.809\pm0.024$$
for NGC\,288, NGC\,362, and NGC\,6218, respectively.

\section*{Acknowledgements}

We acknowledge financial support from the Russian Science Foundation (grant no. 20--72--10052).

We thank the anonymous reviewers for useful comments.
We thank Charles Bonatto for discussion of differential reddening,
Santi Cassisi for providing the valuable BaSTI isochrones and his useful comments,
Aaron Dotter for his comments on DSEP,
Frank Grundahl for providing the Str\"omgren photometric data with their discussion,
Christopher Onken and Sergey Antonov for their help to access the SkyMapper Southern Sky Survey DR3,
Antonio Sollima for providing useful photometric results,
Eugene Vasiliev for his useful comments.
We thank Michal Rozyczka for providing the data for NGC\,6218, which were gathered within the CASE project conducted at the Nicolaus Copernicus
Astronomical Center of the Polish Academy of Sciences.

This research makes use of Filtergraph \citep{filtergraph}, an online data visualization tool developed at Vanderbilt University through
the Vanderbilt Initiative in Data-intensive Astrophysics (VIDA) and the Frist Center for Autism and Innovation
(FCAI, \url{https://filtergraph.com}).
The resources of the Centre de Donn\'ees astronomiques de Strasbourg, Strasbourg, France
(\url{http://cds.u-strasbg.fr}), including the SIMBAD database, the VizieR catalogue access tool and the X-Match service, were widely used in this study.
This work has made use of BaSTI and DSEP web tools.
This work has made use of data from the European Space Agency (ESA) mission {\it Gaia} (\url{https://www.cosmos.esa.int/gaia}), processed by the {\it Gaia}
Data Processing and Analysis Consortium (DPAC, \url{https://www.cosmos.esa.int/web/gaia/dpac/consortium}).
The Gaia archive website is \url{https://archives.esac.esa.int/gaia}.
This study is based on observations made with the NASA/ESA {\it Hubble Space Telescope}.
This publication makes use of data products from the {\it Wide-field Infrared Survey Explorer}, which is a joint project of the University of California, 
Los Angeles, and the Jet Propulsion Laboratory/California Institute of Technology.
This publication makes use of data products from the Pan-STARRS Surveys (PS1).
This study makes use of data products from the SkyMapper Southern Sky Survey. 
SkyMapper is owned and operated by The Australian National University's Research School of Astronomy and Astrophysics. 
The SkyMapper survey data were processed and provided by the SkyMapper Team at ANU. 
The SkyMapper node of the All-Sky Virtual Observatory (ASVO) is hosted at the National Computational Infrastructure (NCI).

\section*{Data availability}

The data underlying this article will be shared on reasonable request to the corresponding author.

\bsp	
\label{lastpage}
\end{document}